\documentclass[preprints,article,accept,moreauthors,LaTeX,dvi2pdf,10pt,a4paper]{Definitions/mdpi} 

\firstpage{1} 
\pubvolume{}
\issuenum{}
\articlenumber{}
\pubyear{2018}
\copyrightyear{2018}
\externaleditor{}
\history{}

\pdfoutput=1

\usepackage{graphicx}
\usepackage{amsmath,amssymb}

\Title{The Vacuum State of Primordial Fluctuations in Hybrid Loop Quantum Cosmology}

\Author{Beatriz Elizaga Navascu\'es $^{1,\dagger}$, Daniel Martín de Blas $^{2,\ddagger,\dagger}$ and Guillermo A. Mena Marug\'an $^{2,}$*$^{,\dagger}$}

\AuthorNames{Firstname Lastname, Firstname Lastname and Firstname Lastname}

\address{%
	$^{1}$ \quad Institute for Quantum Gravity, Friedrich-Alexander University Erlangen-N{\"u}rnberg, Staudstra{\ss}e 7, 91058~Erlangen, Germany; beatriz.b.elizaga@gravity.fau.de  \\
	$^{2}$ \quad Instituto de Estructura de la Materia, IEM-CSIC, Serrano 121, 28006~Madrid, Spain}

\corres{Correspondence: mena@iem.cfmac.csic.es}

\firstnote{These authors contributed equally to this work.} 
\secondnote{E-mail: dmartindeblas@gmail.com}

\abstract{We investigate the role played by the vacuum of the primordial fluctuations in hybrid Loop Quantum Cosmology. We consider scenarios where the inflaton potential is a mass term and the unperturbed quantum geometry is governed by the effective dynamics of Loop Quantum Cosmology. In this situation, the phenomenologically interesting solutions have a preinflationary regime where the kinetic energy of the inflaton dominates over the potential. For these kind of solutions, we~show that the primordial power spectra depend strongly on the choice of vacuum. We~study in detail the case of adiabatic states of low order and the non-oscillating vacuum introduced by Mart\'{\i}n~de~Blas and Olmedo, all imposed at the bounce. The adiabatic spectra are typically suppressed at large scales, and display rapid oscillations with an increase of power at intermediate scales. In the non-oscillating vacuum, there is power suppression for large scales, but the rapid oscillations are absent. We argue that the oscillations are due to the imposition of initial adiabatic conditions in the region of kinetic dominance, and that they would also be present in General Relativity. Finally, we~discuss the sensitivity of our results to changes of the initial time and other data of the model.}



\begin{document}

\section{Introduction}	
	
We have entered an era of increasingly accurate observations for a variety of phenomena that have cosmological origins or are influenced by strong gravitational fields, such as the power spectra of the Cosmic Microwave Background (CMB) \cite{precision1,precision2,planck-inf,planck1,planck2} or the emission of gravitational waves during the coallescence of black hole mergers \cite{waves1,waves2}. These data allow us to try and falsify alternative theories of gravity that may modify the predictions of General Relativity (GR). Among these modifications, those~that are due to quantum effects are especially interesting, because they open new avenues to learn about the consequences of the quantization of the spacetime geometry and elucidate which candidates  are viable for a quantum theory of gravity. One of the most solid candidates is Loop Quantum Gravity (LQG) \cite{lqg}, a non-perturbative canonical quantization of GR that is background independent. In LQG, the variables that encode the geometric information, rather than being constructed directly from the metric induced on the spatial sections and its extrinsic curvature, as in geometrodynamics \cite{wald,giulini}, are~holonomies along loops and fluxes through surfaces, defined in terms of the Ashtekar-Barbero $su(2)$ connection and a densitized spatial triad canonically conjugate to it \cite{lqg}. 
	
The application of the quantization techniques of LQG to cosmological systems is known as Loop Quantum Cosmology (LQC) \cite{ap,lqc1}. LQC has got a notable success in homogeneous and isotropic models of the Friedmann-Lema\^{\i}tre-Robertson-Walker (FLRW) type \cite{ap,aps,iaps,mmo}. In~particular, LQC~predicts that the big bang singularity is avoided by means of repulsive quantum geometry \mbox{effects \cite{iaps,mmo}}. Moreover, there exists an ample family of quantum states in LQC with a highly semiclassical behavior at large volumes such that they remain peaked during all their \mbox{history \cite{ap,taveras}}. The peak determines an effective trajectory with remarkable properties \cite{ap,iaps}. For instance, for~flat spatial topology, it coincides with the GR trajectory except when the matter density becomes considerably large, of the order of, let us say, a few thousandths of the Planck density. In that region of large densities, the trajectory still follows a Hamiltonian dynamics \cite{ap,iaps,ACH}, but this dynamics departs from GR and is generated by an effective Hamiltonian that includes quantum geometry corrections. In agreement with our previous comments, the trajectory is never singular, with the Einsteinian Big Bang replaced with a minimum in the volume that is strictly positive, and that is called the Big Bounce~\cite{iaps}. This bounce occurs when a maximum matter density $\rho_{\rm max}$, named the critical density, is~reached~\cite{ap,iaps}.
	
To analyze more realistic situations in cosmology, homogeneous and isotropic LQC has been extended to include small anisotropies and inohomogeneities, treated as perturbations. This~has been done adopting different approaches \cite{effective0,effective,hybr1,dressed1,edward}. Two of these approaches do not modify the dispersion relations of the perturbations, providing in particular a proper ultraviolet behavior that is compatible~with the observed power spectra. These are the hybrid \cite{hybr1,hybr2,hybrnum, hybr3,hybr4,hybr5,hybrten,hybrpred,hybrgui,hybrferm} and dressed metric approaches \cite{dressed1,dressed2,dressed3,AMorris,Agullo,AAG,ABS}.~Although there are plenty of similarities between these two approaches, the~main distinctions follow from a slightly different strategy for the quantization of the perturbations~\cite{hybrmass}. While the hybrid approach follows a canonical quantization of the whole system formed by the homogeneous geometry and its perturbations, subject to the constraints inherited from GR, the dressed metric approach incorporates the most important quantum effects on the homogeneous background in a dressed geometry and describes the evolution of the perturbations as test fields propagating in this dressed background. In the regime of effective LQC for the homogeneous geometry, and ignoring any backreaction, the two strategies that we have explained result in different time-dependent masses for the equations of motion of the perturbations, when these are written in a harmonic-oscillator form \cite{hybrmass}. The discrepancies between the masses occur, precisely, in the region where the quantum effects are relevant and the effective background differs from an Einsteinian~solution.
	
Using the effective dynamical equations of the background and the modified equations of motion obtained in LQC for the cosmological perturbations, that are valid in the pre-inflationary epoch as well as during inflation, one can compute the cosmological power spectra \cite{hybrpred,hybrgui,dressed3,AMorris,AAG}. To do so, some important pieces of information are still missing: the initial conditions. The conditions for the homogeneous geometry are usually fixed by requiring a good phenomenological fit with the most recent observations. We will comment further on this issue later in our discussion. The~situation with respect to the initial data for the perturbations, on the other hand, is more intrincate. Actually, these~data determine the initial vacuum state of the perturbations. Obviously, the power spectra depend on this state. The fact that the history of the Universe is extended to the preinflationary era in LQC does not only alter the natural choice of this vacuum adopted in standard inflation, but~also has consequences in the evolution of the quantum fluctuations, since no vacuum state is known that remains invariant under the resulting dynamics. These effects change the power spectra with respect to GR and add new features to them. The resulting predictions can be contrasted with the available observations of the CMB \cite{hybrgui,dressed3,AMorris} in order to support or reject the proposed approach to inhomogeneous LQC, combined with the specific choice of vacuum selected for the primordial~perturbations.
	
In this work, we will focus our attention on the hybrid approach to LQC, although most of our comments and results can be generalized without much complication to the case of the dressed metric approach. A fundamental hypothesis in the hybrid approach \cite{hybrgowdy1,hybrgowdy2,hybrgowdy3,hybrgowdymatt,hybrid}, shared by the dressed metric formalism in its basic guidelines, is that the most relevant quantum geometry effects are those that affect the homogeneous geometry, which is therefore analyzed using the characteristic polymeric quantization of LQC, whereas the cosmological perturbations can be described using a more standard, Fock quantization. During the last years, certain criteria have been put forward in the context of hybrid LQC for the unique selection of a family of unitarily equivalent Fock representations for the perturbations \cite{uniqueness1,uniquenessferm}. These criteria consist, basically, in the requirement of vacuum invariance under the spatial isometries of the (FLRW) background geometry and in the existence of a dynamical evolution for the associated creation and annihilation-like variables that can be implemented unitarily \cite{unitarity1,unitarity2,unitarity3,modeunit,unitarity4,unitarity5,bianchi}. Even if these criteria have proven to single out the representation up to unitary equivalence, the choice of a specific vacuum state in the Fock space of this representation is still open. Actually, in the cases studied so far, the Fock representation selected in this way turns out to contain physically appealing candidates for a vacuum such as adiabatic and Hadamard states (see e.g., Ref. \cite{unitarity4}). On the other hand, this choice of vacuum can be further restricted by imposing additional mathematical or physical requirements, for instance a well-defined quantum Hamiltonian or a regularizable backreaction~\cite{hybrfermback}. The determination of this vacuum is of the greatest importance from a conceptual point of view, with~implications for all kind of systems that evolve away from stationarity. In addition, as we have pointed out, its fixation is essential to obtain definite results about the spectra of the cosmological perturbations, and hence to produce predictions that eventually could be confronted with the observations.
	
The rest of the paper is organized as follows. We first review the treatment of cosmological perturbations in hybrid LQC in Section \ref{cosmoperturbations}. In Section \ref{uniqueness} we summarize the criteria used in hybrid LQC to pick out a unique family of unitarily equivalent Fock representations for the quantum fields, and an associated Heisenberg dynamics that is unitarily implementable. Besides, we comment on the freedom available in this family and how to restrict it by demanding extra conditions on the Fock quantization. In Section \ref{data} we comment on the dynamical system obtained for the homogeneous geometry and the cosmological perturbations in hybrid LQC, the specification of initial data for them and, in particular, several proposals that have appeared in the literature to fix the data of the perturbations corresponding to a vacuum. We further discuss the choice of initial conditions and model parameters for the FLRW background within effective LQC in Section \ref{background}, where we also analyze the typical behavior of the most interesting background solutions. Section \ref{vacuaspectra} deals with the computation of the power spectra for several possible choices of initial vacua and the comparison between them, setting the bounce as initial time. In Section \ref{choiceoftime} we investigate which effects on the spectra are genuine quantum geometry modifications. We~also consider changes in the initial time where the vacuum state is fixed. A similar study is carried out in Section \ref{lqcparameters}, but in this case varying a parameter of the model that arises from LQG, namely the Immirzi parameter \cite{immirzi}. Finally, we conclude in Section \ref{conclusions}. Throughout the paper, we set the speed of light and the Planck constant equal to the unit, $c=\hbar=1$.

\section{Cosmological Perturbations in Hybrid LQC}\label{cosmoperturbations}
	
In this section, we introduce the perturbed FLRW cosmology that will be studied in the rest of the work. Let us start with the unperturbed cosmological system.
	
\subsection{Unperturbed Cosmology} 
	
The unperturbed system is an FLRW spacetime, described by a scale factor $a$, with canonical momentum $\pi_a$, and a homogeneous scalar field $\phi$ that renders the system non-trivial and plays the role of the inflaton, with momentum denoted by $\pi_{\phi}$. These two canonical pairs comprise the phase space of this homogeneous and isotropic cosmology (previous to the imposition of the Hamiltonian constraint that reflects the invariance under time reparametrizations and that we will discuss below). The inflaton will be subject to a potential given by a mass term, $V(\phi)=m^2\phi^2/2$. We will treat the mass $m$ as a parameter of the system. In addition, we will focus our attention on the case of flat spatial topology, that we furthermore take to be compact, with spatial sections homeomorphic to a three-torus. The spacetime metric can be written in the form \begin{equation}\label{flrwmetric}
		ds^2 = - N_0^2(t) dt^2 +a^2(t) ~^0 h_{ij} d \theta^i d \theta^j ,
\end{equation}
where $N_0$ is the homogeneous lapse function, $~^0 h_{ij}$ is the Euclidean three-metric, and the angular coordinates $\theta_i$ have period equal to a parameter $l_0$. The Euclidean volume of the spatial sections is therefore ${\cal V}_0=l_0^3$. We will call it the fiducial volume \cite{ap}. The physical volume, on the other hand, is~${\cal V}=a^3{\cal V}_0$. 
	
Instead of the geometrodynamic variables $(a,\pi_a)$, the description in LQC is made in terms of the Ashtekar-Barbero connection and the densitized triad \cite{ap,lqc1}. The physical freedom in these geometric variables is captured by the canonical pair $(v,b/2)$. The variable $v$ is proportional to the physical volume, and hence to the cube of the scale factor, while its canonical momentum $b/2$ is a connection variable adapted to the length of the edges on which the fundamental holonomies of the system are computed according to the so-called improved dynamics prescription, put forward in Ref. \cite{iaps}. Their~relation with $(a,\pi_a)$ is
\begin{equation}
		\label{ABvariables}
		| v |= \frac{l_0^3 a^3}{2 \pi G \gamma \sqrt{\Delta_g}} ,\quad \quad v b = - \frac{2}{3} a \pi_{a}.
\end{equation} 
	
The sign of $v$ determines the orientation of the triad, $\gamma$ is the Immirzi parameter \cite{immirzi}, and $\Delta_g$ is the area gap \cite{ap,iaps}. One usually sets $\gamma = 0.2375$, a value that has been argued to reproduce the Bekenstein-Hawking law for the entropy and area of black holes in LQG \cite{BHlqg1,BHlqg2}. The area gap, on~the other hand, is the minimum positive area eigenvalue allowed in the model according to LQG \cite{lqg}. Since~the homogeneity and isotropy of the model require that, for any intersection with a surface, one has as many entering edges as outcoming ones, the area gap is in fact twice the lowest possible nonzero area eigenvalue, obtained with only one insertion instead of two. One gets $\Delta_g= 4 \pi \sqrt{3} G\gamma $.
	
The energy density $\rho$ and pressure $P$ of the inflaton are, respectively, the sum and the difference of its kinetic energy density $\rho_{\rm kin}$ and its potential $V(\phi)$. Explicitly,
\begin{equation}\label{densitypressure}
		\rho = \frac{\pi_{\phi}^2} {8  \pi^2 G^2 \gamma^2 \Delta_g v^2 }+V(\phi) \quad\quad P=\rho- 2V(\phi),
\end{equation}	
and so $\rho_{\rm kin}=(\rho+P)/2$ and $V(\phi)=(\rho-P)/2=m^2\phi^2/2$. Finally, this homogeneous model is subject to one constraint, $H_{|0}$, that can be regarded as the constraint induced by the homogeneous part of the Hamiltonian constraint of GR:
\begin{equation}\label{effh}
		H_{|0}=\frac{1}{4\pi G \gamma \sqrt{\Delta_g} v }\left[\pi_\phi^2- 3\pi G v^2 b^2 + 4 \pi^2 G^2 \gamma^2 \Delta_g v^ 2 m^2 \phi^2 \right].
\end{equation}
	
\subsection{Hybrid Model}
	
We now introduce perturbations around the homogeneous FLRW cosmology described in the previous subsection. These perturbations affect both the massive scalar field and the spacetime metric. We then truncate the action of the system at quadratic order in the perturbations. In this procedure, the homogeneous cosmology variables, described by zero modes, are treated exactly up to the adopted order of truncation.
	
Using the isometries of the fiducial Euclidean spatial metric and the availability of the Laplace-Beltrami operator associated with this metric, we expand all the perturbations in Fourier harmonics, classified in scalar, vector, and tensor harmonics. The perturbations can be described in terms of the Fourier coefficients of these expansions. From these perturbative variables, the only ones that encode physical information are those that are invariant under perturbative spacetime diffeomorphisms, usually called gauge invariant perturbations \cite{bardeen}, and that in our system are the tensor modes and a canonical pair of scalar field invariants. Since we are considering the case of flat spatial topology, it is convenient to choose the Mukhanov-Sasaki (MS) scalar gauge invariant \cite{mukhanov,sasaki,kodasasa} and a suitable canonical momentum of it \cite{hybr4}.
	
Let us call $T_{\vec{k}}^{(\epsilon)}$ and $Q_{\vec{k}}$, respectively, the coefficients of the tensor and the MS modes in their expansion in terms of harmonics. Here, $\vec{k}$ is the wavevector of the mode, with wavenumber $k$ equal to its (Euclidean) norm. Besides, $\epsilon$ denotes the two possible polarizations of the gravitational tensor modes. Then, a rescaling of these coefficients by the scale factor leads to new mode variables \cite{hybr4} that turn out to satisfy equations of the harmonic-oscillator type \cite{hybrgui,hybrmass}:
\begin{equation}
		\label{ginvariantvariables}
		{\bar d}_{\vec{k},\epsilon}= \frac{a T_{\vec{k}}^{(\epsilon)}}{ \sqrt{32\pi G} \, l_0^{2}} \quad,\quad  {\bar v}_{\vec{k}}= \frac{a Q_{\vec{k}}}{ l_{0}^{2}}.
\end{equation}
	
Actually, it is possible to perform a canonical transformation in all the perturbative variables such that the new canonical set is formed by (combinations of) the linearized perturbative constraints that generate perturbative diffeomorphisms, canonical momenta of them that describe gauge degrees of freedom, the mode coefficients for the gauge invariant perturbations introduced above, and canonically conjugate gauge invariant momenta. 
	
Perhaps even more remarkable is the fact that, at our order of perturbative truncation in the action, the canonical transformation that we have introduced for the perturbations can be extended to a canonical change for the whole system, formed by these perturbations and by the zero modes \cite{hybr4}. The difference between the original and the new zero-mode variables are quadratic contributions in the perturbations. This difference can be regarded as a backreaction effect, and can be ignored in practice if one is interested only in the expression of the metric and of the inflaton up to linear perturbative corrections, or in quantities derived from such expressions. Owing to this and in order to simplify the notation, we will not introduce new symbols for the corrected zero-mode variables, but keep on employing the original ones.
	
With these canonical transformations, the system is described in a especially suitable manner. In particular, it is straightforward to convince oneself that physical states can depend at most on (a~complete set of compatible variables for) the zero modes and the gauge invariant perturbations~\cite{hybr4}. Such states still have to satisfy a global constraint, reminiscent again of the zero mode of the Hamiltonian constraint of GR, and that is given by the sum of a homogeneous contribution that is formally identical to $H_{|0}$ and the Hamiltonians of the gauge invariant perturbations, namely the tensor Hamiltonian, $^{T}H_{|2}$, and the MS Hamiltonian, $^{s}H_{|2}$ \cite{hybr4}. This constraint is represented as an operator, constructed in terms of the basic quantum operators of LQC and those of the Fock quantization selected for the perturbations. 
	
To search for solutions to this quantum constraint, and motivated by the interest in situations where the homogeneous background is not importantly affected by the perturbations, one~often introduces an ansatz for the physical states in which the dependence on the homogeneous geometry and on each of the perturbative modes separates. In this separation, each of the partial wavefunctions of the state is allowed to vary also with the inflaton, that in this way adopts the role of an internal time, in terms of which one can describe the evolution of the rest of variables. Moreover, if one assumes that the presence of the perturbations in the global constraint does not contribute in a significant manner~to produce transitions in the quantum state $\Psi$ of the homogeneous geometry, one arrives at a master constraint for the perturbations by simply taking the expectation value of the original constraint over~$\Psi$~\cite{hybr4}. From this master constraint, it is not complicated to derive equations of motion for the modes of the gauge invariant perturbations (which are basically the counterpart of the Heisenberg equations for the corresponding creation and annihilation operators). As we anticipated, these dynamical equations are of the harmonic-oscillator type. The equations provide the evolution in a conformal time $\eta$ that is obtained with a change of time defined by the expectation value in $\Psi$ of the scale factor, represented in terms of the operator version of $v$. The time-dependent masses in these equations are different for the tensor and the MS gauge invariants, but both are independent of the specific mode under consideration \cite{hybrgui}. These masses are given by ratios of expectation values of geometric operators. By means of these expectation values, the mode equations capture the quantum effects on the FLRW geometry and incorporate them on the dynamical evolution of the perturbations. 
	
In fact, if one follows instead the dressed metric approach, one arrives to a rather similar result. In~that case one also starts with the homogeneous system, quantizing it in LQC. Then, one captures the most relevant part of the LQC effects in a dressed metric defined by a reduced number of expectation values on the geometry. The dynamics of this dressed metric is lifted to the truncated phase space that includes the gauge invariant perturbations, ignoring any backreaction. In this way, one obtains evolution equations for the perturbations in which all geometric terms are replaced with functions of the basic expectation values that characterize the dressed metric. In particular, if one adopts the mode variables ${\bar d}_{\vec{k},\epsilon}$ and ${\bar v}_{\vec{k}}$ as before, their dynamics is governed by equations of the harmonic-oscillator type, with time-dependent masses that are just functions of the already mentioned expectation \mbox{values \cite{dressed2,dressed3,hybrmass}}.
	
\subsection{Effective LQC Regime}
	
A physically interesting regime is found if the quantum state $\Psi$ of the homogeneous geometry in the hybrid approach (or the considered dressed metric in the alternative approach) is a solution of the FLRW model that follows a trajectory of the effective LQC dynamics. In particular, it follows that in this case no backreaction is contemplated (the self-consistency of this procedure has to be checked in practice in the dressed metric scenario \cite{dressed2,dressed3}). In this regime, the expectation values that determine the time-dependent masses of the mode equations for the perturbations, as well as the definition of conformal time, can be approximated by their evaluation on the effective trajectory of the peak of $\Psi$. This trajectory is generated by the effective Hamiltonian constraint
\begin{equation}\label{effH0}
		H_{|0}^{\rm eff}= \frac{1}{4 \pi G \gamma \sqrt{\Delta_g} v }[\pi_{\phi}^2- 3 \pi G v^2 \sin^2{b} + 4 \pi^2 G^2 \gamma^2 \Delta_g v^{2} m^2 \phi^2],
\end{equation}
which is the counterpart of $H_{|0}$ obtained with the replacement of the classical term $b^2$ with $\sin^2{b}$. This~replacement is due to the fact that the connection variable $b$ is not well defined as an operator in LQC, and has to be represented in terms of holonomies, the basic elements of which are imaginary exponentials of $b/2$. This, together with the reality of the $b$-contribution to the constraint and the need to recover its classical expression in the limit of a small connection variable, are at the root of the appearance of a square sinus in $H_{|0}^{\rm eff}$. 
	
The effective equations for the inflaton are the same as in GR, because the part of the Hamiltonian constraint corresponding to the inflaton is the same in GR and in effective LQC. Nevertheless, the evolution equations of the geometry change. The Friedmann and Raychaudhuri equations of GR, that in terms of the Hubble parameter $H$ can be written in the form $H^2=8\pi  G\rho/3$ and $\dot{H}=-4\pi G(\rho+P)$, with the dot denoting the derivative with respect to the proper time, become now in the effective regime \cite{ap}:
\begin{equation}
		\label{effectiveHubble}
		H^2=\frac{8\pi  G}{3} \rho \left( 1 -\frac{\rho}{\rho_{\rm max}}\right), \quad \quad \dot{H}=-4\pi G(\rho+P)\left( 1 -2 \frac{\rho}{\rho_{\rm max}}\right).
\end{equation}
	
Recall that $\rho_{\rm max}=3/(8 \pi G \gamma^2 \Delta_g)$ is the critical density. In particular, it is straightforward to see that the Big Bounce, where the Hubble parameter vanishes (and hence the physical volume reaches a minimum), occurs when the energy density $\rho$ equals $\rho_{\rm max}$. Besides, $\rho$ cannot exceed $\rho_{\rm max}$ (at least in this effective description), because otherwise the square of the Hubble parameter would be negative according to Equation \eqref{effectiveHubble}, something that is not possible.
	
Within this effective regime, the time-dependent masses of the gauge invariant modes are geometric quantities evaluated on the effective trajectories, and the mode equations in hybrid LQC~become
\begin{eqnarray}\label{tensorGR}
		{\bar d}_{ \vec{k},\epsilon}^{\prime\prime} + \left[  {\bar k}^2+ \left( \frac{2\pi G v}{\gamma^2 \Delta_g} \right)^{\frac{2}{3}} \left( \sin^2{b} - 4\pi G \gamma^2 \Delta_g   m^2 \phi^2 \right)  \right] {\bar d}_{\vec{k},\epsilon}&=& 0, \\
		\label{MSGR}
		{\bar v}_{ \vec{k}}^{\prime\prime}+ \left[{\bar k}^2+ \left( \frac{2\pi G v}{\gamma^2 \Delta_g} \right)^{\frac{2}{3}} \left( \sin^2{b} - 4\pi G \gamma^2 \Delta_g   m^2 \phi^2 \right)  + {\bar U}_{\rm MS} \right] {\bar v}_{\vec{k}}&=& 0,
\end{eqnarray}
where ${\bar k}=l_0 k$ is the rescaled wavenumber independent of the fiducial volume, and the prime denotes derivative with respect to the conformal time $\eta$ introduced above, related with the proper time $t$ within this effective regime by $l_0a(\eta)d\eta=dt$ \cite{continuum}.  Besides, the extra term in the equation of the scalar modes, that we will call the MS potential, is 
\begin{equation}
		\label{hybrMSpotential}
		{ \bar{U}_{\rm MS} }=\left( 2 \pi G \gamma\sqrt{\Delta_g} v\right)^{\frac{2}{3}} \left[m^2+ \frac{4  \pi_{\phi}} {v} \frac{\sin(2b)} {\sin^2 b } m^2 \phi +24 \pi G  m^2 \phi^2- \frac{32 \pi^2 G^2 \gamma^2 \Delta_g} { \sin^{2} b} m^4 \phi^4 \right].
\end{equation}
	
We can now specify exactly our definition of time-dependent mass: we will call so the $k$-independent term that multiplies the gauge invariant variable in the above equations. We will denote the corresponding tensor and MS masses, respectively, by $s^{(t)}$ and $s^{(s)}$. In particular, we have that $s^{(s)}-s^{(t)}={ \bar{U}_{\rm MS} }$.
	
Comparing Equations \eqref{tensorGR}--\eqref{hybrMSpotential} with those obtained for the same gauge invariant modes in GR~\cite{continuum}, we notice the replacement of $b^2$ with $\sin^2{b}$ in the part of the time-dependent masses that is independent of the MS potential, and the change in the expression of this potential with respect to the classical one:
\begin{equation}
		\label{Mspotential}
		U_{\rm MS}=\left( 2 \pi G \gamma\sqrt{\Delta_g} v\right)^{\frac{2}{3}}  \left[m^2 +8 \frac{\pi_{\phi}} {vb} m^2 \phi+24 \pi G m^2 \phi^2 - \frac{32 \pi^2 G^2 \gamma^2 \Delta_g } {b^2} m^4 \phi^4  \right]. 
\end{equation}
	
These equations hold in the continuum Fourier limit for the mode spectrum of the perturbations. This limit can be reached by extracting a reference scale from the physical volume, e.g., the value at the bounce (or at present, if preferred), and letting this scale grow unboundedly \cite{continuum}. Obviously, by construction, the resulting system is independent of the original reference scale, which loses all physical significance.
	
The mode equations in the dressed metric approach are also similar in this effective regime, but~with changes in the time-dependent masses of the tensor and MS perturbations, owing to the different quantization strategies. Consequently, the discrepancies between the two approaches appear only in the regions where the quantum effects are important. For a detailed discussion about these differences, the reader can consult Ref. \cite{hybrmass}.  
	
\section{Selection of the Fock Quantization}\label{uniqueness}
	
As commented in the introduction, a key feature for the robustness of the hybrid approach is the characterization of a unique family of unitarily equivalent Fock representations for the description of the cosmological perturbations. This specification is attained by imposing the invariance of the representation under the spatial isometries of the flat FLRW cosmology and by asking a non-trivial and unitarily implementable Heisenberg dynamics of the annihilation and creation operators, when~considered as evolving on a classical background \cite{unitarity5}. Actually, the imposition of these conditions, that we will summarize below, has been successful in selecting a unique equivalence class of states not only for background, homogeneous and isotropic cosmologies with flat compact spatial sections, but also in the case of constant non-zero curvature \cite{uniqueness1,unitarity1,unitarity2}.
	
\subsection{Preferred Fock Representations for Scalar and Tensor Perturbations}
	
One can check that, classically, the tensor and scalar perturbations characterized by the mode coefficients ${\bar d}_{\vec{k},\epsilon}$ and ${\bar v}_{\vec{k}}$ satisfy equations of motion, in conformal time, equivalent to those of the modes of a Klein-Gordon field on a Minkowski spacetime with a time-dependent mass given by $s^{(t)}$ and $s^{(s)}$, respectively. In particular, these mode equations dynamically decouple from each other, and they only depend on the wavenumber $k$, but not in the rest of information contained in the wavevector $\vec{k}$ of the mode. This feature can be traced back to the isometries of the homogeneous and isotropic spatial slices. In fact, it can be argued \cite{unitarity3} that any choice of annihilation and creation-like variables that defines a vacuum state which is invariant under the action of those isometries must be constructed as follows. For each mode, the annihilation-like variable must be a linear combination of the configuration variable ${\bar d}_{\vec{k},\epsilon}$, or ${\bar v}_{\vec{k}}$, and of its canonical momentum, with coefficients that only depend on the wavenumber $k$. The creation-like variable is then the complex conjugate. Let us call $f_k$ and $g_k$, respectively, the coefficients of the configuration and momentum variables, without making any distinction in the notation for tensor and scalar perturbations as they can be analyzed in a parallel way. These coefficients are not completely arbitrary, because they are constrained by the requirement that the resulting annihilation and creation-like variables have Poisson brackets equal to $-i$.
	
Since the evolution of the tensor and scalar mode variables is a canonical transformation, the~annihilation and creation-like variables evaluated at different instants of time are related by Bogoliubov transformations. From a standard quantum mechanical point of view, one would then say that the Heisenberg dynamics of the annihilation and creation operators is unitarily implementable if those Bogoliubov transformations may be represented as unitary operators in the Fock space defined at an arbitrarily given time. This is known to happen if the antilinear part of those transformations is a trace-class operator. The criterion put forward for the choice of a Fock quantization in hybrid LQC is then to determine a family of annihilation and creation-like variables such that their dynamics is unitarily implementable. It is clear that, if $f_k$ and $g_k$ are chosen to be constant, the considered Bogoliubov transformations encode exactly all the dynamical evolution of the tensor and scalar modes ${\bar d}_{\vec{k},\epsilon}$ and ${\bar v}_{\vec{k}}$. However, taking into account that the non-stationary background can be regarded as a dynamical entity on its own (as it should be in the context of quantum cosmology), in principle there exists the possibility of allowing that $f_k$ and $g_k$ vary on time through a dependence on the homogeneous, background variables. Notice that, by doing so, one could in particular consider annihilation and creation-like variables that carried all the evolution of the original mode variables $T_{\vec{k}}^{(\epsilon)}$ and $Q_{\vec{k}}$. Remarkably, the imposition of unitarity on the Heisenberg dynamics associated with the resulting annihilation and creation-like variables turns out to restrict $f_k$ (and consequently $g_k$) in such a way that, in the asymptotic regime of infinite $k$, its dominant part is constant (apart from an arbitrary phase), and equal to the standard choice for a massless Klein-Gordon field on a Minkowski \mbox{spacetime \cite{unitarity1,unitarity2,modeunit,bianchi}}. The subdominant contributions are subject to some convergence conditions in the region of infinitely large wavenumbers $k$. Furthermore, any two families of annihilation and creation-like variables constructed under these requirements of isometry invariance and unitarity can be shown to be unitarily equivalent. The uniqueness of the Fock quantization follows directly from this fact.
	
It should be noticed that this uniqueness criterion allows one to univocally disentangle (at least to dominant order in the asymptotic regime of large wavenumbers) the homogeneous dynamical degrees of freedom, described by means of LQC methods, from the quantum excitations of the perturbations that evolve unitarily when considered on a classical background. This is so thanks to the asymptotic characterization of the background dependence of $f_k$ and $g_k$, coefficients that in turn define the annihilation and creation-like variables chosen for the Fock quantization of the perturbations.
	
\subsection{Further Specification of Vacua}
	
Even though the criterion of a unitarily implementable Heisenberg dynamics, together with the invariance under the spatial isometries, picks out a unique equivalence class of states for the tensor and scalar perturbations, the physical selection of one of these states as the vacuum remains an open issue. As we will explain in the next section, this ambiguity can be equivalently encoded in the choice of initial data for the perturbations, among all the possibilities that are still allowed by the proposed criterion. Actually, this freedom allows us to change how the subdominant contributions of $f_k$ and $g_k$ (in the asymptotic regime of large wavenumbers) depend on the homogeneous background variables. A~physically appealing possibility is to use this freedom, and hence remove (at least part of) the ambiguity, looking for annihilation and creation-like variables such that their Hamiltonian admits a quantum representation with good properties. In fact, there already exist some preliminary investigations about this issue in the technically simpler case of fermionic perturbations around flat FLRW spacetimes, within the hybrid LQC approach \cite{hybrferm,hybrfermback}. For a Dirac field in this cosmological scenario, one can also carry out a characterization of the annihilation and creation-like variables that define invariant vacua under the symmetry transformations of the system and display a unitarily implementable dynamics when considered on a classical background, all this while retaining a standard convention about the respective identification of particles and antiparticles \cite{uniquenessferm}. Similarly to the situation described above for the tensor and scalar perturbations, the proposed criterion picks out an equivalence class of states for the fermionic perturbations, and allows one to separate in an essentially unique form the degrees of freedom of the homogeneous background from the quantum fermionic excitations (in the asymptotic regime of large wavenumbers). The permitted families of annihilation and creation-like variables are still affected by a considerable ambiguity, like in the case of the tensor and scalar perturbations. Nonetheless, if one imposes that the Hamiltonian that dictates the dynamics of the annihilation and creation operators in this fermionic case be well-defined on the vacuum, and therefore densely defined in Fock space, it is possible to prove that one arrives at further restrictions on the time dependence (through the homogeneous background variables) of the subdominant contributions of the coefficients that are analogous to $f_k$ and $g_k$ for the Dirac field. Moreover, these restrictions turn out to guarantee that certain contributions of the fermionic backreaction that affect the quantum evolution of the homogeneous geometry are well defined and finite without the need for any regularization \cite{hybrfermback}.
	
\section{Dynamical System and Initial Conditions}\label{data}
	
The dynamical equations of the perturbed cosmological system that we introduced in Section~\ref{cosmoperturbations} can be integrated numerically to determine the evolution of the different variables during the preinflationary and inflationary eras. This integration requires the specification of initial data, both for the homogeneous sector and for the gauge invariant perturbations. In this section, we are going to propose and discuss possible choices for these data.
	
\subsection{Initial Data}
	
We will consider first the homogeneous part of our system. Let us notice that we can use the freedom to remove a reference scale, commented above, and set in this way, e.g., the initial physical volume equal to one. In addition, since no backreaction of the perturbations is contemplated, the~homogeneous variables must satisfy the effective Hamiltonian constraint $H_{|0}^{\rm eff}$, that can be employed at the initial instant of time, for instance, to find the initial value of the inflaton momentum, and hence of the time derivative of the inflaton on the effective solution. Moreover, the Big Bounce provides a privileged spatial section of minimum physical volume. Therefore, the bounce appears as a natural choice of initial time, where the time derivative of the physical volume vanishes. In this manner, we~are left only with one initial piece of data for the homogeneous, background effective solutions: the~initial value of the inflaton.  We can also view as a phenomenological parameter the inflaton mass, the~value of which can be adjusted, together with that of the inflaton at the bounce, in order to obtain a good fit of the CMB power spectra, although still allowing for the existence of quantum effects. These~effects will typically appear in the large scale sector of the spectra, since these are the scales that may be affected by quantum changes in the Hubble parameter around the bounce if the observed Universe inflated from a region of, approximately, the Planck size. The explained adjustment leads to values of the inflaton at the bounce, $\phi_B$, slightly smaller than one, and to masses of the order of $10^{-6}$, both in Planck units \cite{hybrgui} (see also Ref. \cite{AMorris} in the case of the dressed metric approach). For definiteness, in the rest of our discussion we will perform all of our numerical integrations taking the values $m=1.2\times 10^{-6}$ and $\phi_B=0.97$.
	
Let us now analyze the issue of the initial data for the rescaled perturbations ${\bar d}_{\vec{k},\epsilon}$ and ${\bar v}_{ \vec{k}}$. We notice that fixing these data amounts to determining a complete set of mode solutions for the perturbations. In~fact, we can choose the initial conditions so that they only depend on the wavenumber, $k$, but~not on the particular wavevector corresponding to it, nor on the polarization in the case of tensor perturbations. The complex conjugate of the resulting solutions, including their spatial dependence in terms of Fourier harmonics, are then also dynamical solutions, because the mode equations are real, and the complex conjugation operation leads to harmonics with the same wavenumber. Moreover, we can fix our initial data so that the mode solutions are normalized with respect to the (rescaled) Klein-Gordon inner product. Let us call $\{\mu_k(\eta)\}$ the solutions obtained in this way, together with their complex conjugates $\{\mu_k^{\ast}(\eta)\}$. In our notation, we do not make distinction between the solutions for the scalar modes and for the tensor modes, although it is clear that the equations of motion that they have to satisfy are different. This type of notation will help simplify the discussion. Continuing with our analysis, it is not difficult to check that the Klein-Gordon normalization condition becomes
\begin{equation}
		\label{kleingordon}
		{\mu}_{ k}(\eta_0) {\mu}_{ k}^{\ast\prime}(\eta_0) - {\mu}_{ k}^{\prime}(\eta_0) {\mu}_{k}^{\ast}(\eta_0)=i.
\end{equation}
	
Here, $\eta_0$ is the initial time (thus, $\eta_0$ is the time when the rescaled physical volume reaches a minimum if the initial section is taken at the bounce). 
	
The gauge invariant fields can be expressed in terms of our set of mode solutions (with the spatial dependence given by the Fourier harmonics) as linear combinations with coefficients that are creation and annihilation-like, given our normalization condition \eqref{kleingordon}. Explicitly, ${\bar v}_{\vec{k}}(\eta)=\mu_k(\eta) a_{\vec{k}} +\mu_k^{\ast}(\eta) a_{\vec{k}}^{\ast}$, and similarly for the tensor modes. Therefore, fixing initial data for the perturbations amounts to describing the fields in terms of a specific collection of creation and annihilation-like coefficients. Besides, the choice of these coefficients determines a unique Fock representation, in which the creation and annihilation operators are the direct counterpart of our coefficients. Furthermore, since the vacuum state is the state of unit norm characterized by a vanishing action of all the annihilation operators of the Fock representation, the specification of these latter operators picks out a unique vacuum for the system. In this way, we see that providing initial data for the perturbations is tantamount to the choice of a vacuum state for the gauge invariant tensor and MS fields. Moreover, our construction of the normalized perturbative solutions in terms of Fourier harmonics (of the fiducial Laplace-Beltrami operator) determines a particular type of vacua that have the good property of being invariant under the spatial isometries of the fiducial spatial metric \cite{AGvacio1}.  
	
We can always absorb a global phase in the definition of our mode solutions without any relevant consequence. The creation and annihilation-like variables will only change in a constant phase, and~therefore the associated annihilation operators will still have a vanishing action on the same vacuum state. We use this freedom to set the initial value of the mode solutions equal to a positive number. In~this manner, we can parametrize the initial data for these solutions as \cite{hybrgui}
\begin{equation}
		\label{parametrization}
		{\mu}_{k}(\eta_0) = \frac{1}{\sqrt{2D_{k}}}, \quad \quad {\mu}_{ k}^{\prime} (\eta_0)= \sqrt{\frac{D_{k}}{2}}\left(C_{k}-i\right).
\end{equation}
	
Here, we have employed that the imaginary part of ${\mu}_{ k}^{\prime} (\eta_0)$ is determined by ${\mu}_{ k}(\eta_0) $ and the normalization condition \eqref{kleingordon}. The constant $D_k$ is positive, while $C_k$ can take any real value. 
	
\subsection{Adiabatic Vacuum States}
	
Possible choices of a vacuum for the gauge invariant fields are given by the so-called adiabatic states \cite{adiabatic,adiabaticLR}. These states were originally introduced as approximated solutions to the equations of fields propagating in cosmological spacetimes and have a particularly well-tamed ultraviolet behavior and, associated with that, good regularization properties (at least for adiabatic states of sufficiently high order \cite{adiabaticreg1,adiabaticreg2}). 
	
The adiabatic solutions are based on an ansatz of the form: 
\begin{equation}
		\label{adiabaticsol}
		{\mu}_{ k} (\eta)= \frac{1}{\sqrt{2W_{k}(\eta)}}e^{-i\int^{\eta}_{\eta_0}W_{k}(\bar{\eta})d\,\bar{\eta}}.
\end{equation}
	
Then, the mode equations are fulfilled provided that the function $W_k$ satisfies the equation
\begin{equation}
		\label{adiab-sol-W}
		W^{2}_{k} = k^{2} + s - \frac{1}{2}\frac{W^{\prime\prime}_{k}}{W_{k}}+\frac{3}{4}\left(\frac{W^{\prime}_{k}}{W_{k}}\right)^{2},
\end{equation}
where $s$ denotes generically the scalar $(s^{(s)})$ or tensor $(s^{(t)})$ time-dependent mass. Approximate solutions to this equation can be obtained by an iterative process, starting with the function \mbox{$W_k^{(0)}=k$}. At the step $(n+1)$, one substitutes in the right-hand side of Equation \eqref{adiab-sol-W} the approximate solution $W_k^{(2n)}$ estimated at the previous step, obtaining a better approximation, that is called $W_k^{(2n+2)}$. The~approximate iterative solutions $W_k^{(2n)}$ converge to the exact solution at least as $\mathcal{O}(k^{-1-2n})$ in the asymptotic limit of infinite $k$ \cite{hybrgui}, where the symbol $\mathcal{O}$ stands for asymptotic order. Adiabatic states of, say, order $2n$ are then constructed by choosing as solutions ${\mu}_{ k} (\eta)$ those that are uniquely characterized by the initial data $W_k^{(2n)}$ at time $\eta_0$, at least for sufficiently large $k$. The asymptotic behavior of the resulting solutions restricts severely the form of the adiabatic states in the ultraviolet region, but however leaves full freedom for the behavior at large scales, i.e., at small $k$.
	
This freedom is reflected in the infinite ambiguity that exists in choosing adiabatic states. Its~definition clearly depends on the order of adiabatic iteration in their construction, but also on the initial conformal time $\eta_0$ chosen for the initial data $W_k^{(2n)}$. Moreover, since only the asymptotic behavior for large $k$ is relevant in their definition, we can truncate the iterative solutions $W_k^{(2n)}$ at order $k^{-1-2n}$, obtaining new, but valid approximate solutions $\mathfrak{W}^{(2n)}_{k}$. We will call the states constructed out of these new approximate solutions the obvious adiabatic states. For the sake of an example, at the first non-trivial adiabatic order (i.e., order 2) we get
\begin{equation}
		\label{firstorder}
		W^{(2)}_{k} = \sqrt{k^{2}+s},\qquad \mathfrak{W}^{(2)}_{k} = k + \frac{s}{2k}.
\end{equation}
	
Notice that the adiabatic solutions coincide at lowest, 0th order: 
$ W^{(0)}_{k} = \mathfrak{W}^{(0)}_{k} = k$.
	
As we have commented, adiabatic states depend on the initial time selected to define them. Equivalently, they change with the conformal time chosen to set their initial data. It is straightforward to see, from the adiabatic ansatz, that the parameters that determine these initial data are
\begin{align}\label{adiabaticparameters}
		D_{k} = W_{k}(\eta_0),\qquad C_{k} = -\frac{W^{\prime}_{k}(\eta_0)}{{2W^{2}_{k}(\eta_0)}}.
\end{align} 
	
This parametrization is valid for any adiabatic order, and both for the iterative states and their obvious states counterpart.
	
The correct definition of the adiabatic states needs that the function $W_k$ (or rather the considered approximation) be real. For sufficiently small $k$, this requires the positivity of the time-dependent mass $s$. Remarkably, in the case of hybrid LQC and for the kind of background effective solutions that we are interested in studying, this positivity is granted at the bounce \cite{hybrmass}.
	
Turning back to the freedom inherent to the choice of an adiabatic vacuum, we can impose additional criteria to remove it. For instance, in Ref. \cite{AANvacio} it has been proposed that one could select the state that provides a regularized stress-energy tensor that vanishes mode by mode at the given initial time. Although there is an infinite ambiguity in the adiabatic renormalization process, once one fixes this process (as it is done in Ref. \cite{AANvacio}), the proposal picks out a single adiabatic vacuum for each selection of initial time. Then, the specified vacuum turns out to be unique if one fixes this initial time~univocally.
	
\subsection{Other Vacuum States}
	
A different proposal for the vacuum state of the perturbations has been put forward in \mbox{Refs. \cite{AGvacio1,AGvacio2}}. This proposal fixes the vacuum by requiring a subtle interplay between the behavior in the region with important LQC effects, quantified by an energy density $\rho\geq 10^{-4}\cdot \rho_{\rm max}$, and the behavior at the end of inflation. In the region with relevant quantum effects, the requirement is that the vacuum belong to the ball of states with quantum Weyl curvature below a specific bound. It can be considered the lowest viable bound allowed by the uncertainty principle and stable under quantum evolution throughout all of the studied region. At  the end of inflation, the demand, specialized in detail in Ref. \cite{AGvacio2} to scalar perturbations, is that the state minimizes the square of the Ricci tensor of the spatial metric. It has been shown that, (at least) in the dressed metric approach, this proposal leads to very appealing primordial power spectra, in good agreement with observations \cite{AGvacio2}. Nonetheless, the~direct relation of this vacuum with adiabatic states, or their determination in terms of specific initial data for the perturbations has not been explored.
	
Another proposal for a vacuum state is the so-called non-oscillating vacuum (NO-vacuum) introduced in Ref. \cite{hybrpred}. The criterion to select this vacuum is based on a specific choice of initial data such that the oscillations in the power spectrum are minimized during the evolution in a given time interval. Since the primordial power spectrum for each mode is proportional to the square norm of the mode solution, the criterion can be stated as a variational problem: the determination of the functions $D_{k}$ and $C_{k}$ that minimize the integral
\begin{equation}\label{IO}
		\int_{\eta_{0}}^{\eta_{f}}\left|\frac{d\big( |\mu_{k}|^{2}\big)}{d\bar{\eta}}\right|d\bar{\eta}.
\end{equation}
	
Again, we see that this proposal is not enunciated directly as a condition on the section of initial time, but involves an entire interval $(\eta_0,\eta_f)$. In fact, the selected state will depend in general on the choice of this interval. In the considered case, a natural choice is the period from the bounce to the instant at which the kinetic energy density of the inflaton vanishes, moment at which the inflation is already driven by the potential. We will discuss the sensitivity to changes of this interval later in~Section~\ref{lqcparameters}. 
	
In some specific situations, it is possible to calculate analytically the data $D_{k}$ and $C_{k}$ that minimize the integral \eqref{IO} but, in general, numerical methods will be necessary. A case in which one can carry out an analytical computation is a de Sitter spacetime with a massless scalar field. In this case, if one lets the initial time tend to minus infinity, one can prove that the initial data that minimize the integral reproduce the Bunch-Davies vacuum. In addition, for a scalar field in flat spacetime, either massless or with a quadratic potential, the proposal picks out the  Poincar\'e vacuum state. Actually, in both of these particular situations, it is possible to analytically minimize the oscillations in the power spectrum by choosing a basis of solutions such that the associated annihilation and creation-like variables, when~viewed as operators in a Heisenberg picture, are adiabatic invariants up to a phase (for details and the definition of adiabatic invariants, see Ref. \cite{adinv}). This line of attack implies the resolution of a generalized Ermakov-Pinney equation. The procedure leads as well to the Bunch-Davies and Poincar\'e vacua in the two considered cases, respectively. Given this coincidence, it would be interesting to compare the commented procedure with the NO-vacuum in other more general scenarios where one could attain an explicit construction of adiabatic invariants.
	
The consequences of the choice of the NO-vacuum in hybrid quantum cosmology have been investigated in some detail, both for scalar \cite{hybrpred} and tensor perturbations \cite{hybrgui}. In fact, the resulting power spectra fit particularly well the observational data. 
	
\subsection{Power Spectra}
	
The primordial spectra can be easily calculated if we find the value of the mode solutions at the end of inflation, $\eta_{\rm end}$, e.g., by integration of the mode equations with the initial data provided by some choice of vacuum. The explicit formulas for the scalar and tensor primordial power spectra, $\mathcal{P_{\mathcal{R}}}(k)$ and $\mathcal{P_{\mathcal{T}}}(k)$ respectively, if we assume the same vacuum for both types of perturbations, are the following:
\begin{equation}
		\label{primordialpower}
		\mathcal{P_{\mathcal{R}}}(k)=\frac{k^3}{2\pi^2}  \frac{|\mu_k (\eta_{\rm end})|^2} {z^2(\eta_{\rm end})}, \quad \quad   \mathcal{P_{\mathcal{T}}}(k)=\frac{32 k^3}{\pi}  \frac{|\mu_k (\eta_{\rm end})|^2} {a^2(\eta_{\rm end})},
\end{equation}
where $z=a\dot{\phi}/H$, with the Hubble parameter defined in the standard manner, $H=\dot{a}/a$. These primordial spectra provide the initial conditions for the Boltzmann codes that can be used to compute the CMB angular spectra. We recall that, once we have set the physical volume ${\cal V}$ equal to one at a certain fixed time, the scale factor $a$ can be obtained as the cubic root of ${\cal V}$, with the value of this factor identified as the unit scale at the chosen time (for instance, the bounce). Besides, in practice one can stop the integration at times sufficiently later than the horizon crossing for the evaluation that is needed in the above formulas, as far as one has reached the regime in which the evolution of the mode, suitably rescaled by $z$ (scalar modes) or $a$ (tensor modes), gets frozen. We also recall that a mode with wavenumber $k$ is said to cross the horizon at time $\eta_{\rm cross}$ if we have that $k=a(\eta_{\rm cross})H(\eta_{\rm cross})$.
	
Actually, once the primordial spectra are calculated for a certain vacuum state, we do not need to repeat the computation of those spectra if we change of vacuum in our discussion. As we have seen, each of the vacua is characterized by a set of mode solutions. Let them be $\{ \mu_k(\eta) \}$ and $\{ {\bar \mu}_k(\eta) \}$ for the two vacua in question. The two sets are always related by a Bogoliubov transformation of the form
\begin{equation}\label{bogoliubov}
		{\bar \mu}_{k}(\eta)=\alpha_k \mu_k (\eta )+\beta_k \mu_k^{\ast} ( \eta ) ,
		\quad\quad | \alpha_k |^2-| \beta_k |^2  =1 . 
\end{equation}
	
The last identity in this equation is a consequence of the fact that the Bogoliubov transformation is canonical. The Bogoliubov coefficients can be calculated using the initial data of the two sets of mode solutions:
\begin{eqnarray}
		\label{alfa}
		\alpha_k &=& i \left[ {\bar \mu}_k^{\prime}(\eta_0)   {\mu}_k^{\ast}(\eta_0)-{\bar \mu}_k(\eta_0)   {\mu}_k^{\ast\prime}(\eta_0) \right] , \\ \label{beta}
		\beta_k &=&-i \left[ {\bar \mu}_k^{\prime}(\eta_0)   {\mu}_k(\eta_0)-{\bar \mu}_k(\eta_0)   {\mu}_k^{\prime}(\eta_0) \right].
\end{eqnarray}
	
Then, if we have at our disposal the primordial power spectra $\mathcal{P_{\mathcal{R}}}(k)$ and $\mathcal{P_{\mathcal{T}}}(k)$ of one of the analyzed vacua, let us say the vacuum with mode solutions  $\{ \mu_k(\eta) \}$, the spectra for the other vacuum can be obtained as
\begin{equation}
		{\bar{\mathcal{P}}}_{\mathcal{R}} (k)= \left[1 + 2|\beta_k|^2 + 2|\alpha_k||\beta_k|\cos\left(\varphi^\alpha_{k}-\varphi^\beta_{k} + 2\varphi^{\mu}_{k}\right)\right] \mathcal{P_{\mathcal{R}}}(k), 
\end{equation}
and  similarly for the tensor spectrum, where we have introduced the notation 
\begin{equation}\label{powerchange}
		\alpha_k = |\alpha_k|e^{i\varphi^\alpha_{k}}, \quad\quad \beta_k = |\beta_k|e^{i\varphi^\beta_{k}}, \quad\quad \mu_k = |\mu_k|e^{i\varphi^{\mu}_{k}}.
\end{equation} 
	
This simplifies the numerical computations enormously.
	
As an aside, let us point out a useful property of the Bogoliubov transformations between adiabatic states. For these kind of states, the asymptotic behavior of the $\beta$ coefficients is determined by the vacuum of lower adiabatic order. Let us call $2m$ this order, with our convention of using even integers for the order of the adiabatic states. Then, explicitly, one gets \cite{hybrgui} 
\begin{equation}
		\label{betaadiabatic}
		|\beta_k|=\mathcal{O}(k^{-2-2m})
\end{equation} 
for asymptotically large $k$.
	
Finally, in scenarios where the observed modes cross the horizon during a period of slow-roll inflation (see e.g., Ref. \cite{slowroll}), the primordial power spectrum can be evaluated by using analytical expressions that extend the results from one reference wavenumber, $k_{\rm ref}$, for which the power has been obtained by other means, typically numerically or observationally. This reference mode must exit the horizon also in the slow-roll regime. This regime is reached when the Hubble-flow functions  $\epsilon_{H} = -\dot{H}/H^{2}$ and $\eta_{H}=\ddot{H}/(2\dot{H}H)$ are sufficiently small. For our numerical computations \cite{hybrgui}, we~quantify this condition by requiring that the two parameters are smaller than $10^{-2}$. In this slow-roll regime, the Planck Collaboration uses the formulas \cite{planck-inf}
\begin{eqnarray}\label{slowrollscalar}
		\mathcal{P}^{\rm Pl}_{\mathcal{R}}(k;k_{\rm ref})&=&A_s(k_{\rm ref})\left(\frac{k}{k_{\rm ref}}\right)^{n_s(k_{\rm ref})-1+\frac{1}{2}\ln\left(\frac{k}{k_{\rm ref}}\right)\frac{dn_s}{d\,\ln\, k}(k_{\rm ref})},
		\\ \label{slowrolltensor}
		\mathcal{P}^{\rm Pl}_{\mathcal{T}}(k;k_{\rm ref})&=&A_t(k_{\rm ref})\,\left(\frac{k}{k_{\rm ref}}\right)^{n_t(k_{\rm ref})+\frac{1}{2}\ln\left(\frac{k}{k_{\rm ref}}\right)\frac{dn_t}{d\,\ln\, k}(k_{\rm ref})}.
\end{eqnarray}
	
The spectral indices $n_s$ and $n_t$, and the runnings $dn_s/d\,\ln\, k$ and $dn_t/d\,\ln\, k$, are quadratic polynomials of the Hubble-flow functions. Besides, all of them, as well as the coefficients $A_s$ and $A_t$, depend on the reference mode $k_{\rm ref}$. Their explicit expressions are given in Ref. \cite{planck-inf}. 
	
\section{Background Solution and Effects on the Perturbations}\label{background}
	
In this section, we will analyze the main characteristics of the effective solution for the homogeneous sector. We recall that we are interested in solutions that lead to power spectra compatible with the observations, but still retaining some quantum effects in the region of large scales. This~determines a relatively narrow range of values for the initial condition on the inflaton, and~for the value of the mass parameter. For concreteness, we will consider a specific solution in this set, namely, that corresponding to $m=1.2\times 10^{-6}$ and $\phi_B=0.97$ (in Planck units). The properties of other solutions in our sector of interest are qualitatively similar.

\begin{figure}[H]
		\centering
		\includegraphics[width=14 cm]{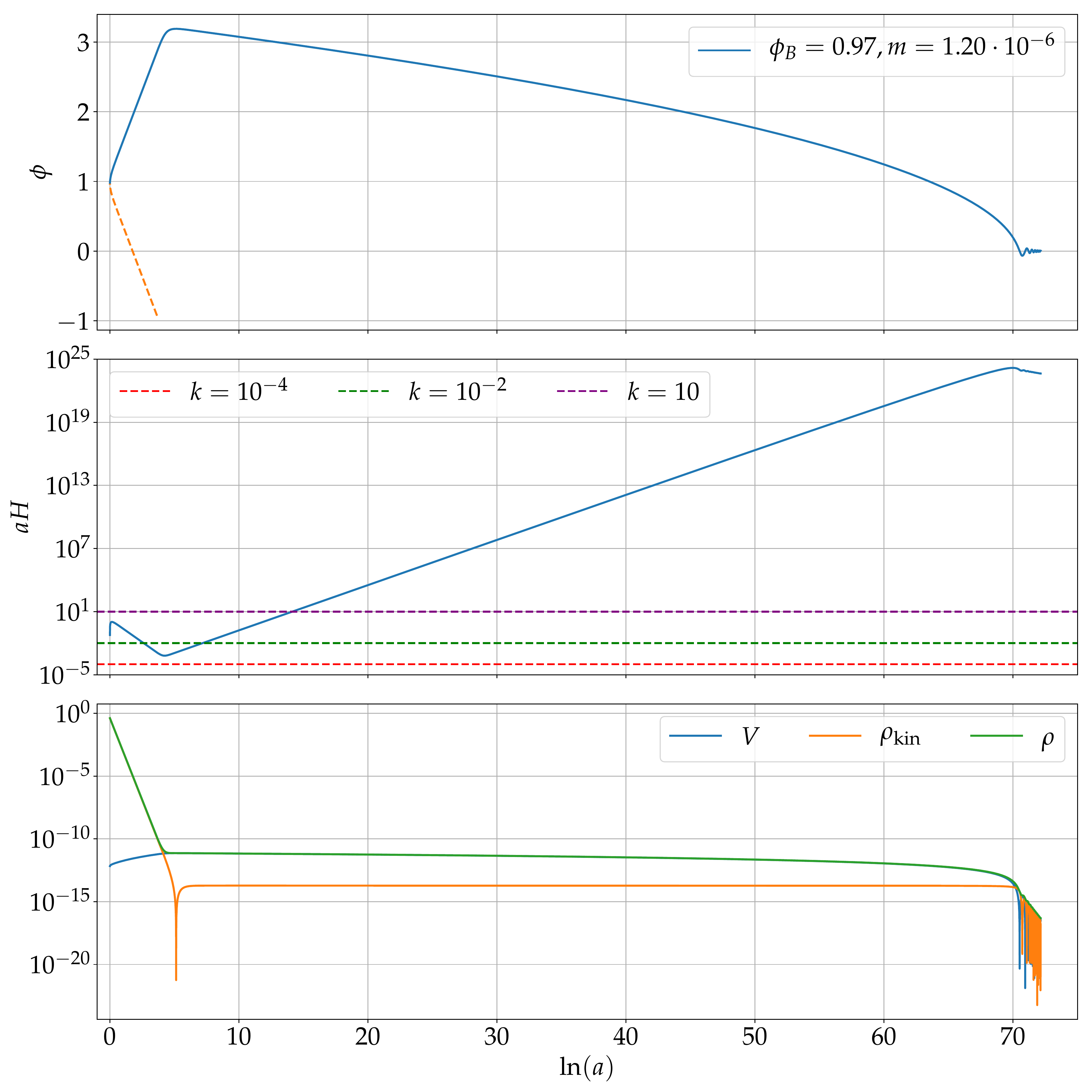}
		\caption{Solution for $m=1.2\times 10^{-6}$ and $\phi_B=0.97$. The upper panel shows the evolution of the inflaton, the middle panel that of the rescaled Hubble parameter $aH$, and the lower pannel that of the energy density, with its kinetic and potential contributions. In the case of $aH$, we have plotted also several wavenumbers to illustrate different scenarios for horizon crossing.}
		\label{fig1} 
\end{figure}

The solution for our choice of $m$ and $\phi_B$ is displayed in Figure~\ref{fig1}. We see that the rescaled Hubble parameter $aH$, that separates the regions of wavenumbers that are inside and outside the Hubble radius, starts at zero at the bounce, grows rapidly in a period of superinflation where the scale factor is almost constant, and then decreases to a minimum. At the bounce, the solution is totally dominated by the kinetic energy of the inflaton, while the contribution of the potential is almost negligible. The~kinetic energy density then decreases rapidly, approximately as $a^{-6} $, as it would correspond to the case with vanishing potential, where the inflaton momentum is conserved and the density evolves as the inverse square of the physical volume. The decrease in the kinetic energy density continues until it becomes of the order of the potential, approximately at the time when $aH$ reaches its minimum, since $\dot{a}$ increases when the potential drives the evolution of the scale factor, as can be seen in Equation \eqref{effectiveHubble}. We note that, during the evolution from the bounce to the minimum of $aH$, the inflaton increases, but only by a small factor of the order of 3. Therefore, the potential changes by one order of magnitude, approximately, during this period. 
	
We recall that the energy density at the bounce is equal to $\rho_{\rm max}$, which is of the order of the Planck density. The energy density when the kinetic contribution becomes comparable to the potential, moment that approximately marks the minimum of $aH$ according to our previous comments, can be estimated as follows. First, notice that the potential at the bounce is $m^2\phi_B^2/2$, which is of the order of $10^{-12}$ in Planck units. Then, it suffices to take into account that, during the phase of decrease of the kinetic energy, the potential grows at most in one order of magnitude. This gives us a potential in the range $[10^{-12},10^{-11}]$. In the case of other effective background solutions in the sector of physical interest, for which $\phi_B$ may be slightly different and its increase in the kinetically dominated regime may be a little bigger, we can safely extend the considerations to the interval of energy densities $\rho_{\rm KE-VE}\in[10^{-12},10^{-9}]$ at the coincidence between the kinetic and the potential contributions. The two densities $\rho_{\rm max}$ and $\rho_{\rm KE-VE}$ play a  fundamental role in the phenomena experienced by the perturbations and, in a certain approximate sense, determine the regions where one can find effects that depart from the standard predictions of GR. 
	
Taking into account Equation \eqref{effectiveHubble} and the fact that the scale factor does not vary much during superinflation, it is not difficult to convince oneself that the maximum of $aH$ close to the bounce is of the order of the Planck unit with our convention of scales. On the other hand, soon after this superinflationary period, the energy density will have decreased sufficiently as to discard any quantum effect in the system, that will get adapted in this way to the GR dynamics. This implies that, from that moment on, $H^2$ starts to decrease as $\rho$, and hence as the inverse square volume, according to our previous comments. Therefore $aH$ decreases as the inverse square of the scale factor. Recalling that the energy density decreases (as $a^{-6}$) during this period by $12-9$ orders of magnitude, we conclude that $aH$ (that goes as $a^{-2}$) decreases to $10^{-4}-10^{-3}$. From all these considerations, it follows that the modes with wavenumbers (a bit) larger than one (in Planck units) do not cross the horizon during these stages previous to inflation, while modes with $1\gtrapprox k\gtrapprox 10^{-4}-10^{-3}$ exit and enter the horizon after the bounce and before inflation. Let us also comment that these modes would be the first to cross the horizon in any later epoch of inflation, therefore noticing any effect that could be attributed to an early stage of the inflationary process. Finally, modes with $k\lessapprox 10^{-4}-10^{-3}$ are not inside the Hubble radius in the immediate vicinity of the bounce and do not reenter the horizon in the studied period.
		
After the potential has become equal to the kinetic energy density, the former starts to dominate while the latter continues to decrease. The scenario is that of a potential that remains essentially constant and drives an epoch of inflationary expansion. From the lower plot in Figure~\ref{fig1} we can see that this period lasts approximately 60--70 e-folds, sufficient to be compatible with the observational bounds \cite{efolds}. Nonetheless, the number of e-folds already suggests an scenario of short-lived inflation. Hence, during the first moments of the inflationary process, the fact that the system is evolving from a kinetically dominated era can lead to departures from a slow-roll behavior. This will have consequences in the power spectra if the modes that crossed the horizon in those instants are observed today (see~e.g., the discussion of Ref. \cite{linde} in the context of GR). In particular, the slow-roll approximation will not be good, at least for modes that exited the horizon during those first stages of inflation. 
	
We close this section by mentioning that, noticing that the potential is practically negligible around the bounce in the kind of background solutions of interest in LQC, the behavior of these solutions has been characterized by approximate analytic expressions in Refs. \cite{waco1,waco2,waco3}, where different potentials have been considered, including the case of a mass contribution. In particular, these analytic expressions are useful to derive general properties of the solutions.
	
\section{Power Spectra for Different Vacua}\label{vacuaspectra}
	
We will now analyze the primordial power spectra of the perturbations around the background described in the previous section, using the initial conditions at the bounce that correspond to different choices of vacuum state. We will consider the case of the scalar power spectrum. The results for the tensor perturbations are similar, and can be found in Ref. \cite{hybrgui}.
	
\begin{figure}[H]
		\centering
		\includegraphics[width=14 cm]{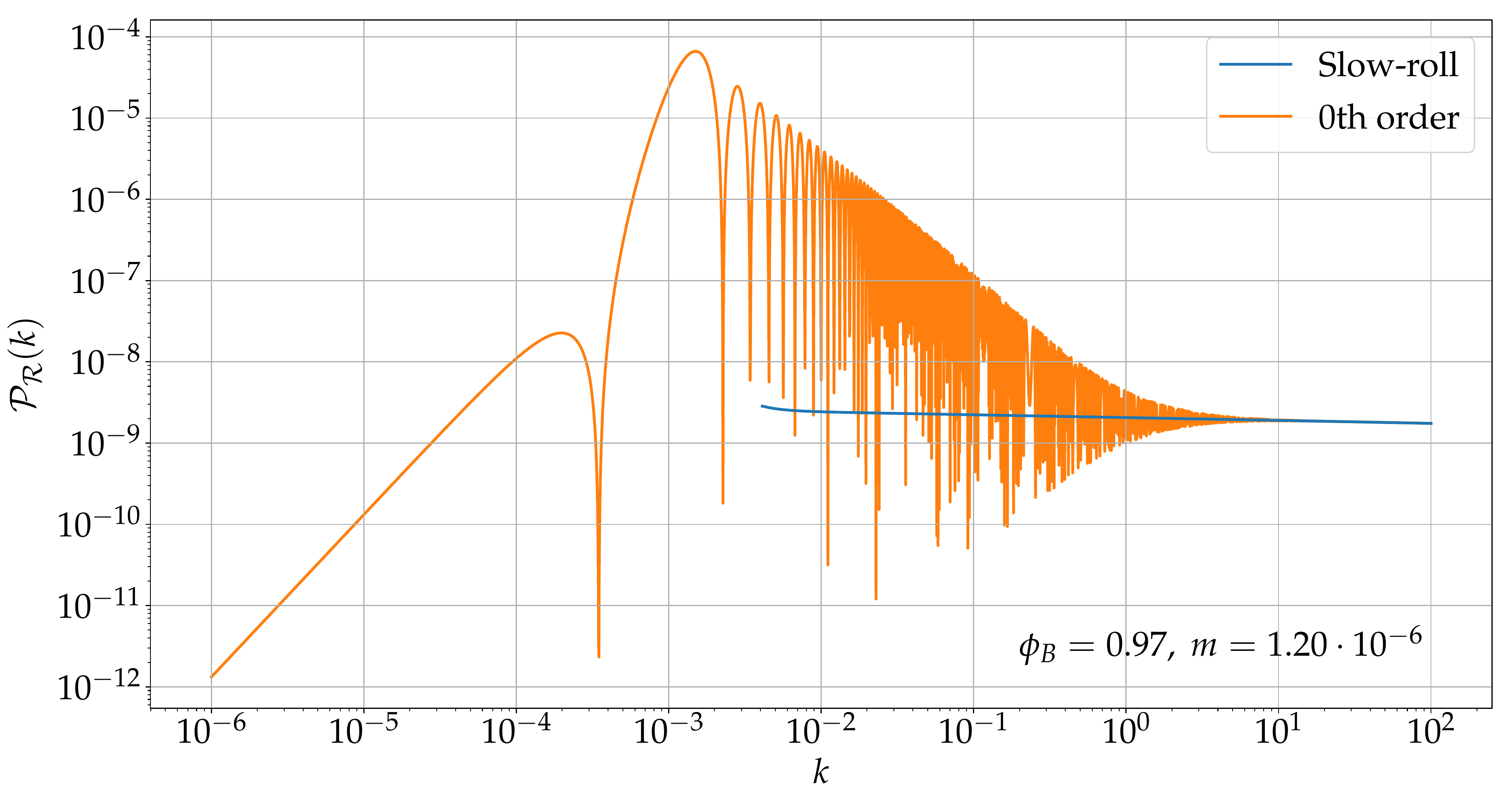}
		\caption{The primordial scalar power spectrum in hybrid LQC for the 0th-order adiabatic state on the background solution studied in Section \ref{background}. The power spectrum obtained with the slow-roll approximation \eqref{slowrollscalar} is also shown for comparison.}
		\label{fig2}
\end{figure}

\begin{figure}[H]
		\centering
		\includegraphics[width=14 cm]{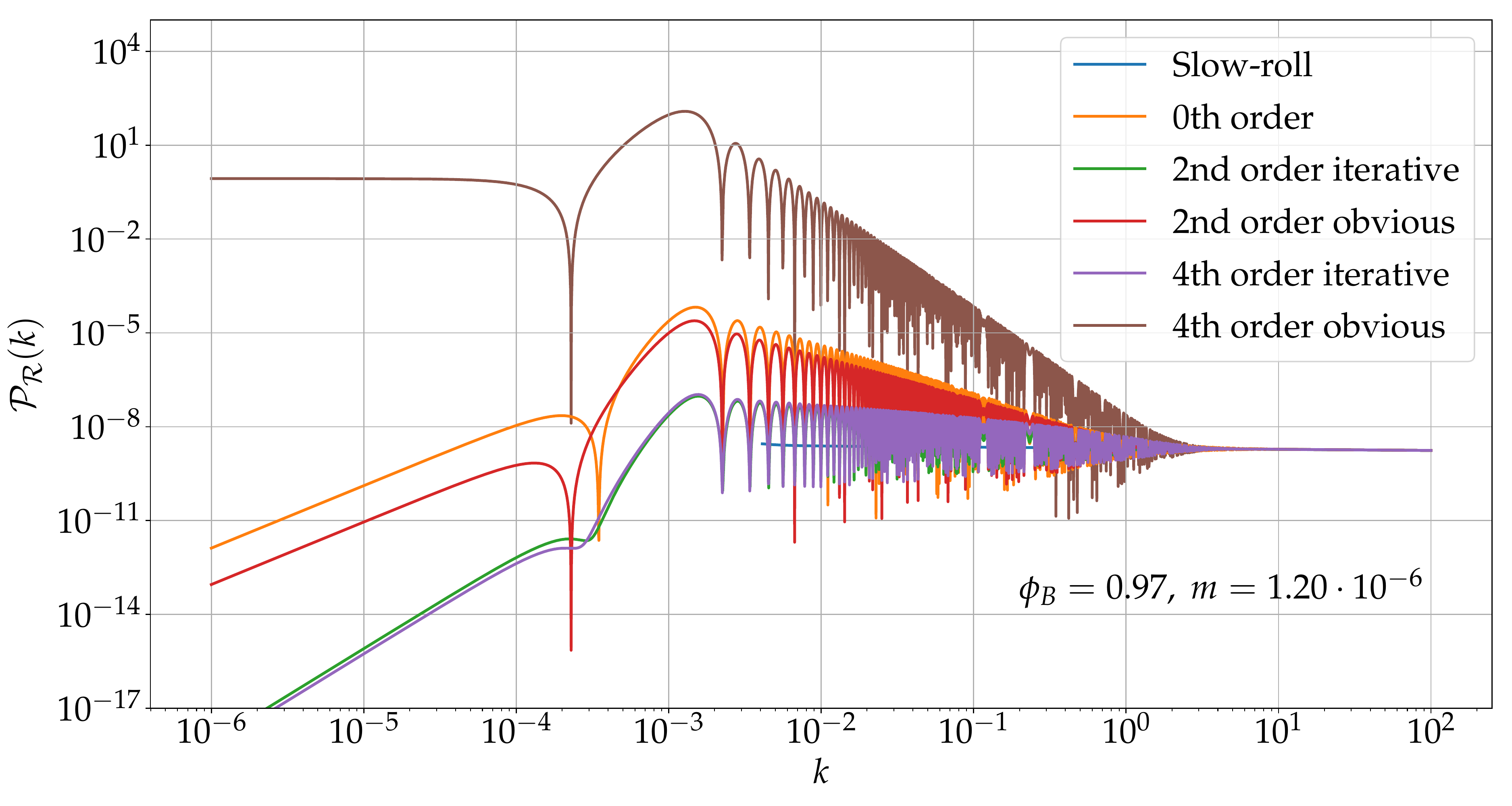}
		\caption{The primordial scalar power spectrum in hybrid LQC for the adiabatic states of lowest orders on the background solution studied in Section \ref{background}. Two kind of adiabatic states are considered: the obvious ones and the iterative adiabatic states. The power spectrum obtained with the slow-roll approximation \eqref{slowrollscalar} is also shown for comparison.}
		\label{fig3}
\end{figure}

In Figures \ref{fig2} and \ref{fig3} we show the scalar power spectra for the lowest order adiabatic vacuum states. The results are compared with the spectrum obtained in the slow-roll approximation with the analytic estimation made by the Planck collaboration, given in Equation \eqref{slowrollscalar}.
	
A look at the power spectrum of the 0th-order adiabatic state in Figure~\ref{fig2} reveals the existence of three regions. First, in the region of wavenumbers $k$ sufficiently greater than one, let us say  $k\gtrapprox 5 $, the~power spectrum is constant and coincides with the slow-roll prediction. Second, in an intermediate region with $5\gtrapprox k\gtrapprox 10^{-4}-10^{-3}$, the spectrum presents rapid oscillations, that grow in amplitude as $k$ decreases. Third, and finally, for $k\lessapprox 10^{-4}-10^{-3}$ the power is suppressed. Notice that there is a clear relation between these three regions and those identified in the previous section according to the behavior of the rescaled Hubble parameter $aH$ on the background solution. This relation lies at the root of the phenomena observed in the power spectrum.  
	
The spectra of Figure~\ref{fig3} show similar results, also with three regions, but we see that the quantitative result depends strongly on the specific vacuum state under consideration. In particular, the suppression of power at large scales is present for all iterative adiabatic states, but not for some of the obvious ones, indicating that the predictions for this latter type of states are less robust against a change in the iterative order. Despite these discrepancies, we emphasize that the regions with different behaviors are the same in all cases.
\vspace{-12pt}

\begin{figure}[H]
		\centering
		\includegraphics[width=13 cm]{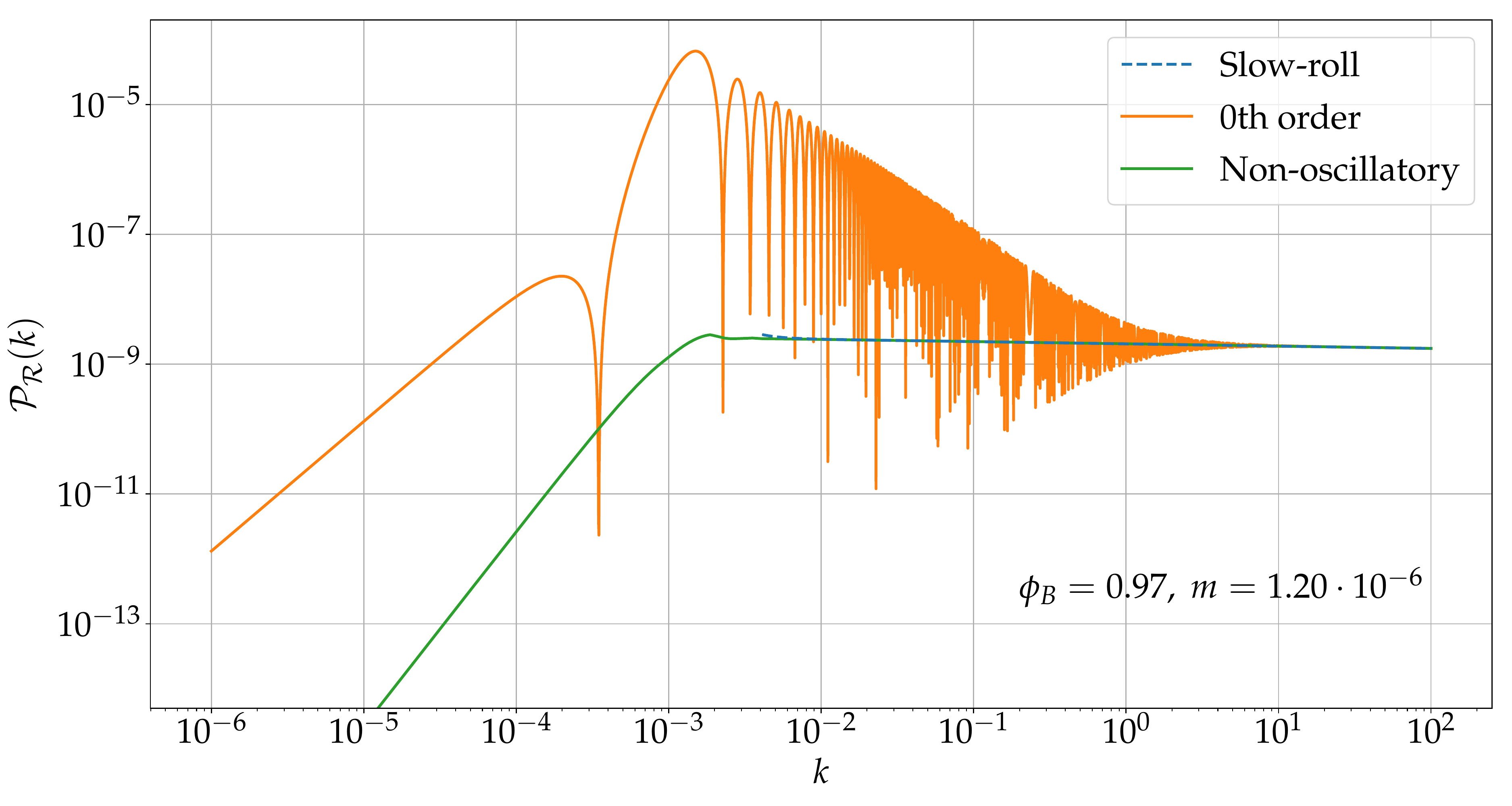}
		\caption{The primordial scalar power spectrum in hybrid LQC for the NO-vacuum on the background solution studied in Section \ref{background}. The power spectrum obtained with the 0th-order adiabatic state and with the slow-roll approximation \eqref{slowrollscalar} are also shown for comparison.}
		\label{fig4}
\end{figure}

The power spectrum for the NO-vacuum is displayed in Figure~\ref{fig4}, comparing it with the corresponding spectra of the 0th-order adiabatic state and the slow-roll approximation. In this case, we find a remarkable coincidence with the slow-roll spectrum in a much wider range of modes, with a total absence of rapid oscillations. The region of small wavenumbers still presents a suppression of power. Finally, we also notice some minor features in the spectrum, in the matching region with the predictions of the slow-roll approximation.

The computation of the angular spectrum of temperature anisotropies in the CMB can be computed using e.g., the \textsf{CLASS} code \cite{class}. We have carried out this integration using the best-fit values of the parameters of the base $\Lambda$CDM model provided by the Planck Collaboration in Ref. \cite{planck-inf} (using the so-called TT+lowP data). The integration has been performed allowing for lensing corrections in the code, in order to improve the accuracy of the results. An additional issue that must be faced before comparing the predictions with the observations is a change in the scale that we have used so far in our calculations of the primordial spectra. This scale was fixed by setting the physical volume at the bounce equal to one. Nevertheless, the observational data usually employ the present value of the physical volume as the unit of reference. This requires a scale matching between the two situations. The matching is made by identifying the scale of reference that the Planck mission uses for $k$, namely the pivot scale $k_{\ast} = 0.05\, \textrm{Mpc}^{-1}$, with the value of $k$ that has the same amplitude observed for \mbox{$k_{\ast}$ \cite{planck-inf}}. The matching is made in the set of modes that crossed the horizon in slow-roll regime, because the spectrum turns out to be monotonous for $k$ in that set, a fact that guarantees that the solution is unique~\cite{hybrgui}. 
	
\begin{figure}[H]
		\centering
		\includegraphics[width=14 cm]{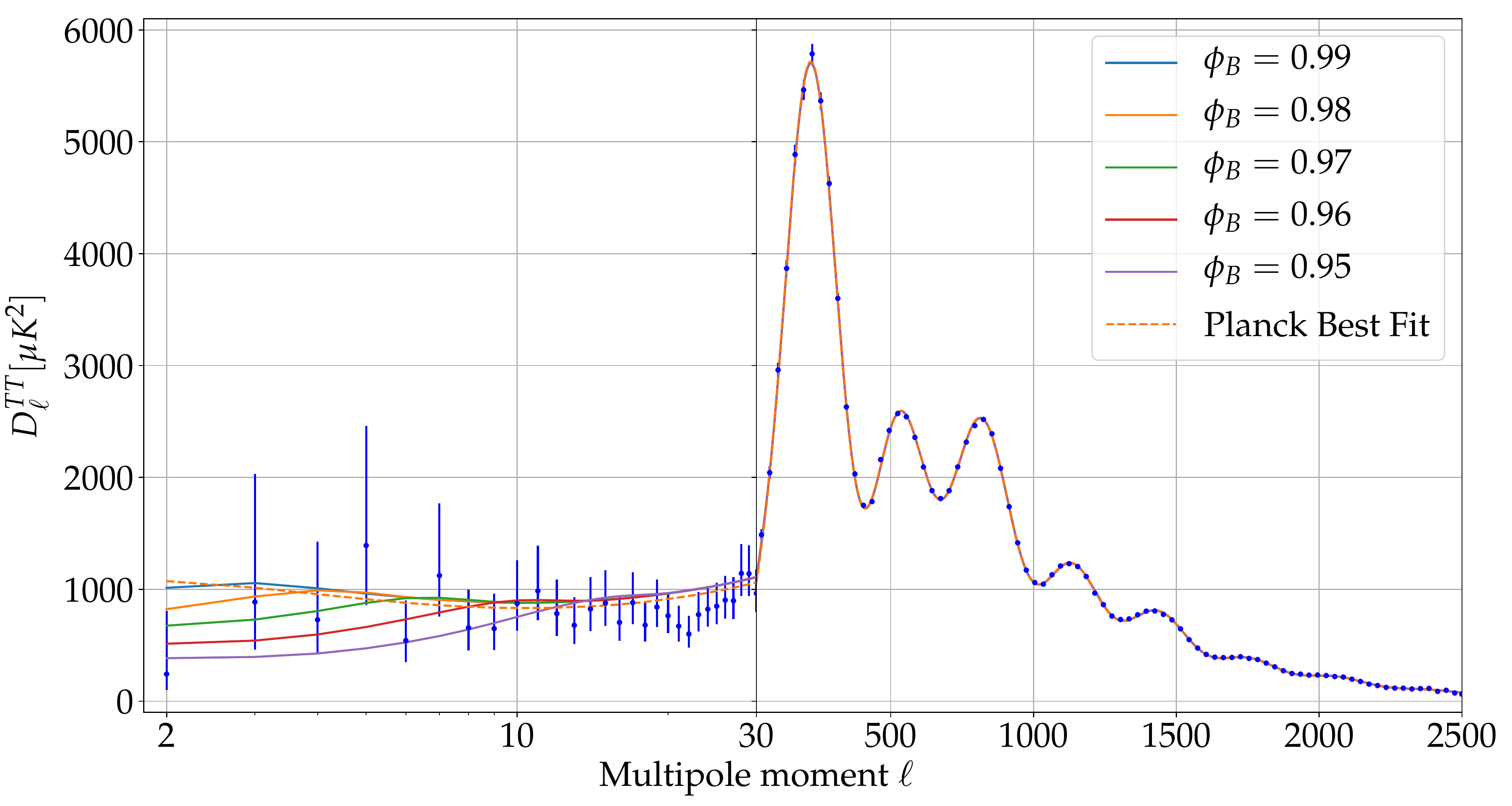}
		\caption{Angular power spectrum of temperature anisotropies for the Planck best fit and the NO-vacuum state at the bounce on background solutions obtained with different values of the inflaton at that bounce and $m=1.20\times 10^{-6}$.}
		\label{fig5}
\end{figure}

The resulting angular spectrum of temperature anisotropies for the NO-vacuum state at the bounce is displayed in Figure~\ref{fig5}. The figure shows the spectrum obtained on the solution studied in Section \ref{background}, but also on other effective background solutions with the same value of the inflaton mass, but slightly different initial data for the inflaton at the bounce. The best fit of the observational data obtained by the Planck Collaboration is displayed as well for comparison. In particular, we notice that the suppression of power and the features that are present in the spectrum of the NO-vacuum may provide an explanation for the lack of power and the possible anomalies that the records of Planck seem to indicate at large and intermediate scales \cite{planck-inf}.
	
\begin{figure}[H]
		\centering
		\includegraphics[width=14 cm]{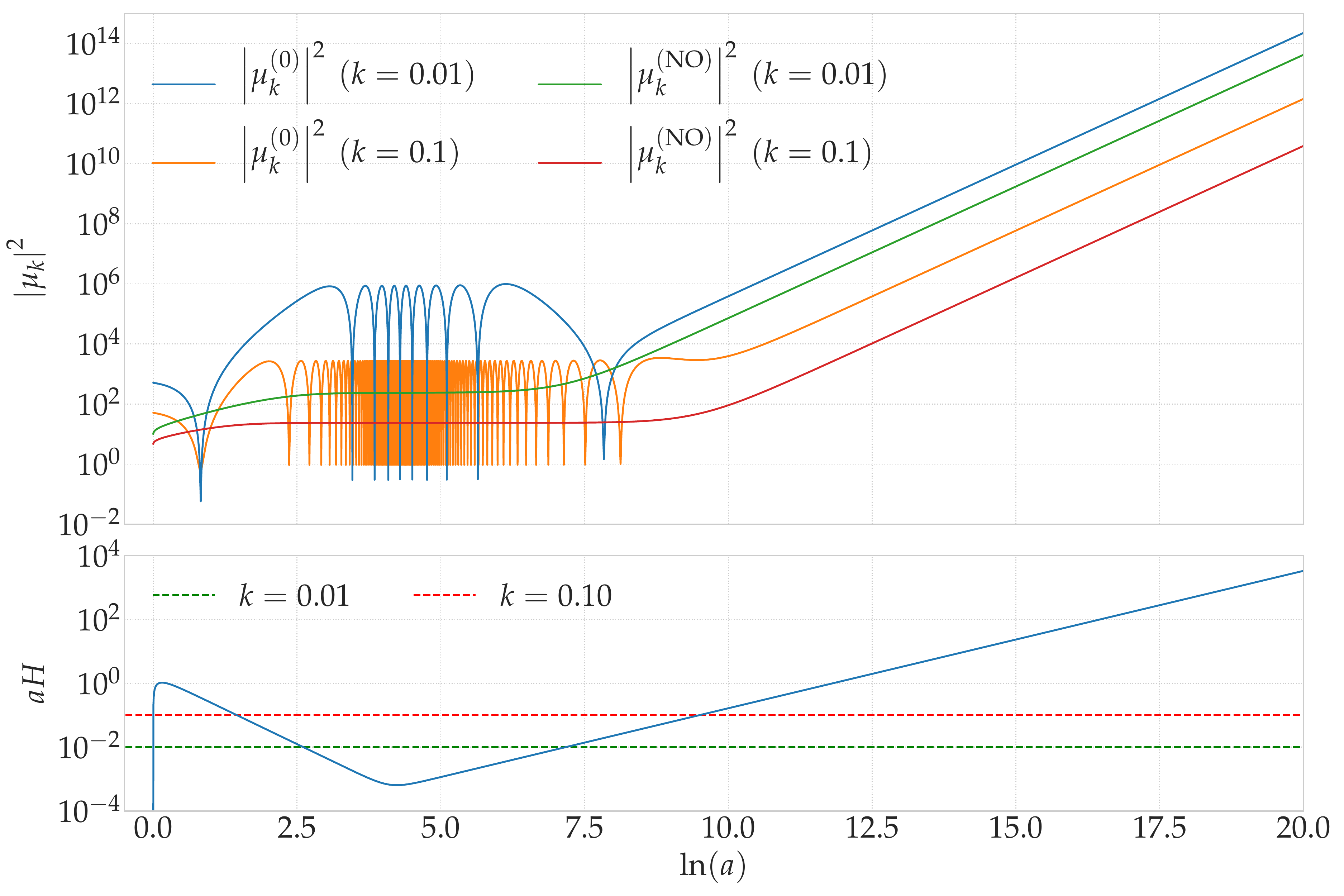}
		\caption{Upper panel: Evolution of two modes in the 0th-order adiabatic state and of the NO-vacuum at the bounce, on the background solution studied in Section \ref{background}. Lower panel: Evolution of $aH$, showing that the two modes exit and reenter the horizon before the onset of inflation, and then cross the horizon again in the first stages of the inflationary period.}
		\label{fig6}
\end{figure}

To help visualizing the origin of these phenomena and make more evident the different behavior of the modes in the distinct vacua, we show in Figure~\ref{fig6} the evolution of two of the modes that exit and enter the horizon in the period between the Big Bounce and the beginning of inflation. We take initial conditions at the bounce that correspond either to the 0th-order adiabatic state or to the NO-vacuum, for comparison. We clearly see a region of rapid oscillations and amplification of the power of the modes in the adiabatic state. This region is related with the period when the modes are inside the horizon, after reentering it in the post-bounce evolution and before they exit again in the early stages of the inflationary process. In the NO-vacuum, on the other hand, the modes do not display these rapid oscillations, but remain almost constant in that region. 
	
\begin{figure}[H]
		\centering
		\includegraphics[width=14 cm]{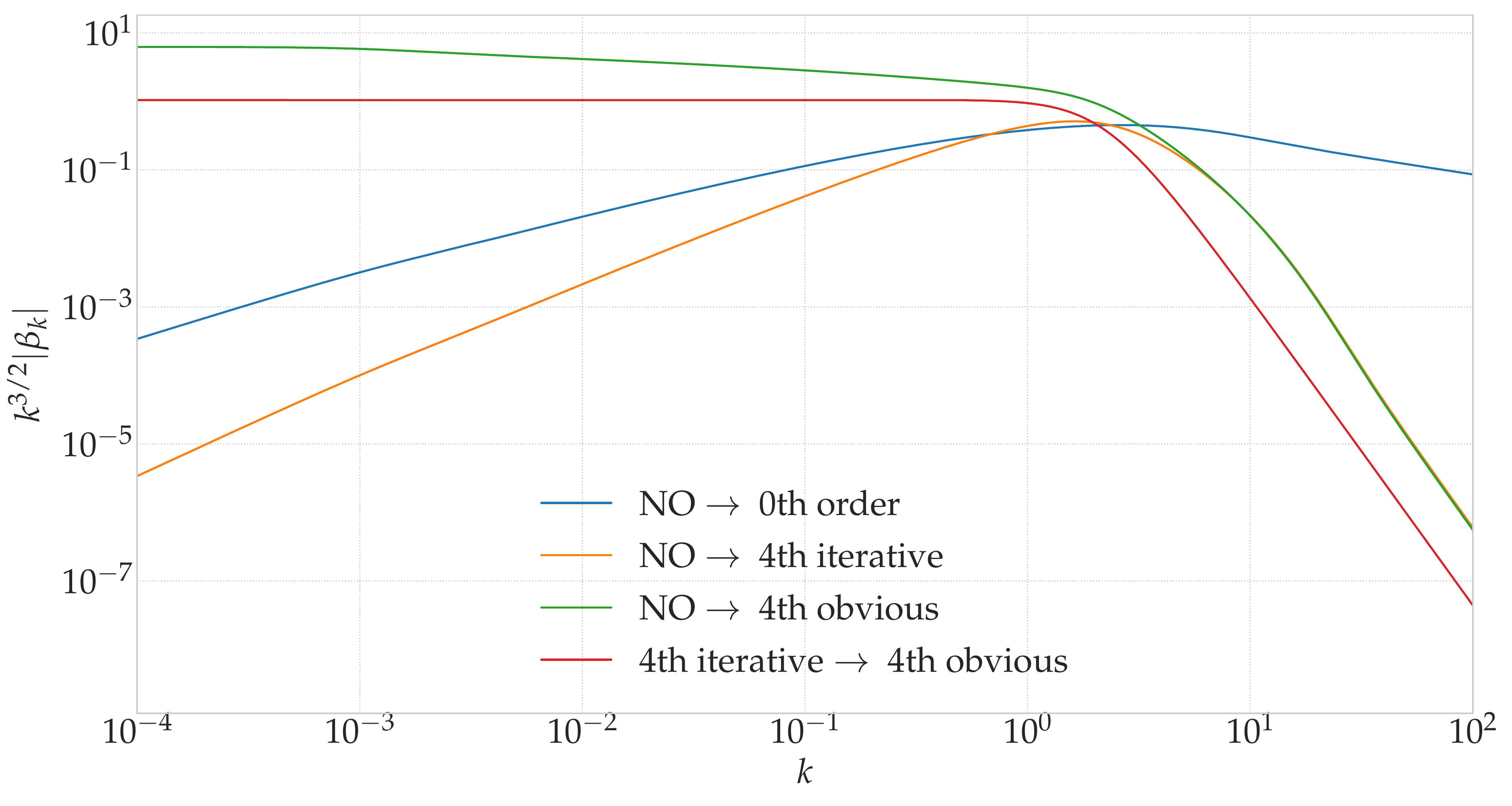}
		\caption{Norm of the beta coefficients, multiplied by $k^{3/2}$, for different Bogoliubov transformations relating adiabatic states of the obvious and iterative types and the NO-vacuum. }
		\label{fig7}
\end{figure}

Finally, Figure~\ref{fig7} shows the beta coefficients for the Bogoliubov transformation between different adiabatic states and the NO-vacuum, as well as between two vacua of 4th order. We consider the two types of adiabatic states introduced in Section \ref{data}, i.e., obvious and iterative adiabatic vacua. According to Equation \eqref{betaadiabatic}, the behavior of the beta coefficient at large $k$ seems to indicate that the NO-vacuum behaves asymptotically as an adiabatic state of high order, at least equal to 4. 
\vspace{-12pt} 
	
\section{Choice of Initial Time and Effect on the Power Spectra}\label{choiceoftime}
	
We have seen that, in general for adiabatic states and always for the iterative ones, one gets a primordial power spectrum that presents a region with rapid oscillations, with the corresponding enhancement of power in comparison with the slow-roll spectrum when these oscillations are averaged. Although one might believe that these rapid oscillations are due to the quantum corrections generated by LQC, a close inspection to the modes that are affected makes us suspect that the oscillations may rather appear because these modes cross the horizon immediately after a period in which the kinetic energy density of the inflaton is important in the background solution. The exposition of the fluctuations to a period of kinetic dominance can have relevant effects in the spectrum \cite{linde}. In~particular, this may put into question the choice of adiabatic sates (at least the lowest order ones) as natural vacua for the perturbations.
	
\begin{figure}[H]
		\centering
		\includegraphics[width=14 cm]{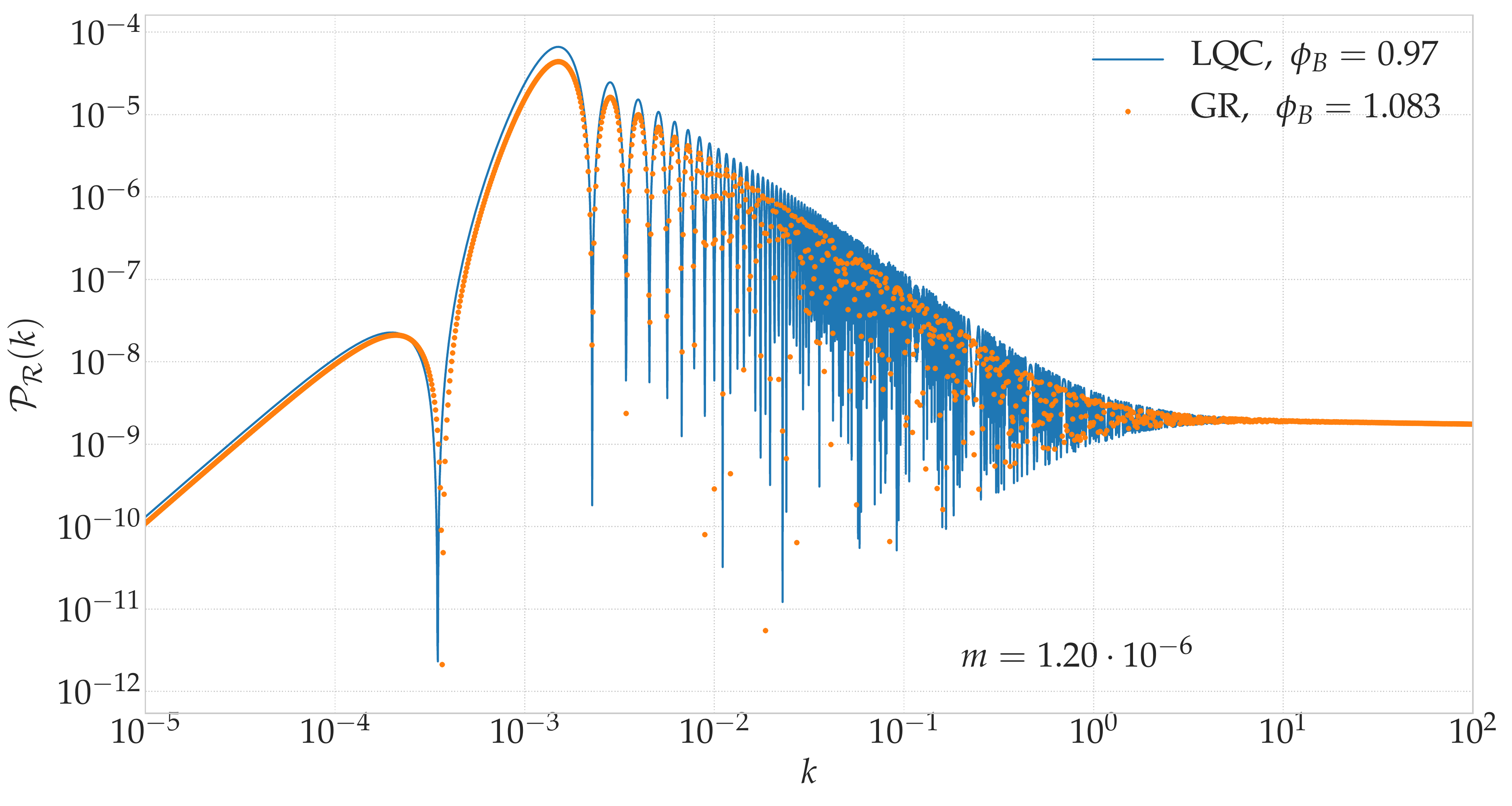}
		\caption{Primordial scalar power spectrum of the 0th-order adiabatic vacuum in hybrid LQC on the solution studied in Section \ref{background}, and in GR on the Einsteinian counterpart of that background. In this background, the inflaton at the time when the bounce occurs in LQC is $\phi_B=1.083$ in Planck units.}
		\label{fig8}
\end{figure}

To understand this issue, we can investigate whether similar oscillations would appear in the primordial spectrum of adiabatic states if we consider background solutions like those in effective LQC, but now in GR. We can try and find solutions in GR that have the same behavior as the original effective ones in the region of large volumes, and continue them back to the regions of high density that correspond to the bounce and its vicinity in LQC. In this way, the GR solutions will coincide with the effective solutions at late times, but not around the initial time, identified with the bounce of the LQC trajectories. In Figure~\ref{fig8} we compare the primordial spectrum of the 0th-order adiabatic state in hybrid LQC on the effective solution considered in Section \ref{background}, with the primordial spectrum of the same state in GR, and on the background constructed also in GR as we have explained above. In~this GR solution (obtained with the same value of the mass), the value of the inflaton at the time that would correspond to the bounce in LQC is $\phi_B=1.083$. We see that the two spectra are similar, with the same kind of oscillations. Furthermore, these oscillations possess similar amplitudes. In this sense, the most relevant departures from the standard inflationary models that are observed in this adiabatic spectrum are not genuinely due to quantum geometry effects, but rather to other phenomena that can be rooted in a kinetically dominated phase previous to the onset of inflation. Obviously, the spectra of hybrid LQC contain also quantum modifications, but in cases like the low-order adiabatic states at the bounce (or in its vicinity), these modifications of the spectra are hidden by others that are related to non-conventional preinflationary and inflationary stages, stages that turn the question of a natural choice of vacuum into an open issue. To analyze in depth the quantum modifications of the spectra, and confront the predictions about them with observations, it is necessary to reach a better characterization of such modifications and disentangle them from the rest of corrections.
	
An important distinction between the possible oscillations in GR and those in LQC is that the latter are confined to a sector of energy densities that is bounded from above by the critical density $\rho_{\max}$, whereas no such a bound exists in the Einsteinian theory.
	
An alternative way to investigate whether the rapid oscillations are an effect of the LQC corrections or are due to other reasons is the following. Until now, we have considered the bounce as a natural choice of time in order to set initial conditions on the state of the perturbations. The final form of this state that determines the primordial power spectrum depends on the evolution of the background solution from the bounce to the end of inflation, and on the specific time-dependent mass that enters the propagation equations. This mass has the same expression as in GR except for corrections that are relevant only in the same region where the effective background solution in LQC differs from its counterpart in GR. In total, we see that all possible quantum corrections, either in the expression of the time-dependent mass or directly in the background solution, are confined to the region with high energy density of the inflaton (let us say greater than $10^{-4}$) that surrounds the bounce. As soon as we get away from the bounce, these quantum effects disappear. Therefore, a procedure to check if certain phenomenon is produced by these effects is to move the initial time away from the bounce, and closer to the onset of inflation. 
	
\begin{figure}[H]
		\centering
		\includegraphics[width=14 cm]{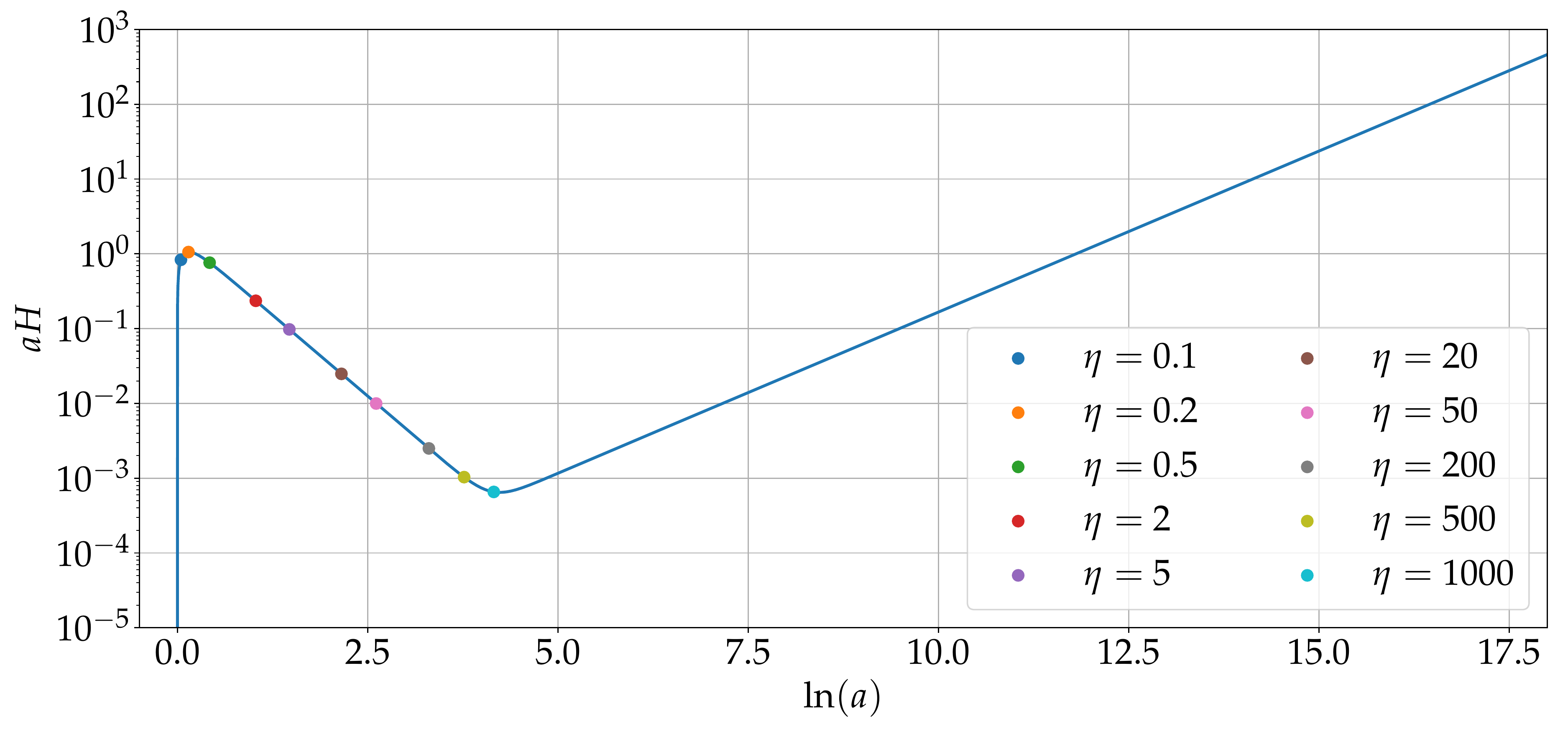}
		\caption{Different choices of time between the bounce and the onset of inflation, and value of $aH$ at those times, for the background solution studied in Section \ref{background}.}
		\label{fig9}
\end{figure}

\begin{figure}[H]
		\centering
		\includegraphics[width=14 cm]{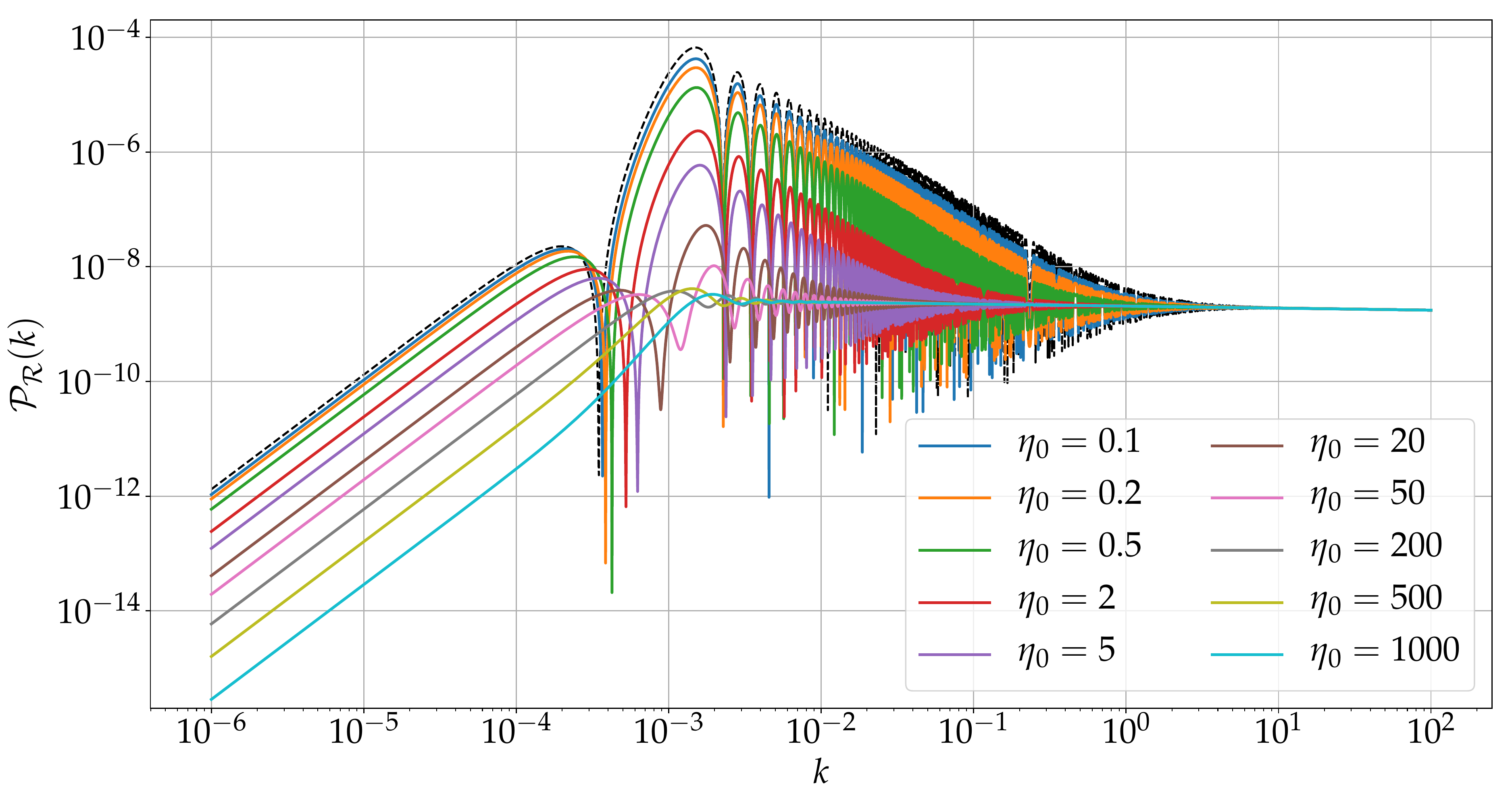}
		\caption{Primordial scalar power spectrum in hybrid LQC of the 0th-order adiabatic vacuum, defined with the different choices of initial time displayed in Figure~\ref{fig9}. The background is the solution studied in Section \ref{background}.}
		\label{fig10}
\end{figure}

In Figure~\ref{fig9} we show a set of alternative choices of time in the interval between the bounce and the onset of inflation, displaying the value that the rescaled Hubble parameter $aH$ reaches at that time in the background solution. We then plot in Figure~\ref{fig10} the primordial power spectrum obtained by taking initial conditions at those times, corresponding to the respective 0th-order adiabatic vacuum state in hybrid LQC on the background solution analyzed in Section \ref{background}. It is clear that the rapid oscillations are removed as the initial time gets closer to the onset of inflation. This is consistent with the statement that these oscillations are not a genuine product of the quantum geometry, but rather of the epoch of kinetic dominance and of its effects on the evolution of the perturbations.
\vspace{-12pt} 
	
\section{Dependence on Parameters}\label{lqcparameters}
	
Our definition of the  NO-vacuum state depends on the choice of a time interval in which the oscillations in the power of the vacuum solution are minimized, mode by mode. Until now, we have chosen the Big Bounce as initial time $\eta_0$ for this interval, and the final time $\eta_f$ has been identified with the moment in which the kinetic energy of the inflation vanishes for the first time, something that happens when the inflationary expansion has already started. Numerical analyses show that the results do not depend heavily on our particular choice of final time if the changes are small. These changes generically affect the features that are present in the spectra at intermediate scales, for wavenumbers $k$ in the region that separates the sector with suppression of power from the sector where the slow-roll approximation applies. As the changes in the final time increase, letting it approach the bounce, the~power spectra start to display oscillations, with the subsequent enhancement of power on average. The closer $\eta_f$ gets to the region with quantum geometry effects, the larger the increase of power. 
	
\begin{figure}[H]
		\centering
		\includegraphics[width=14cm]{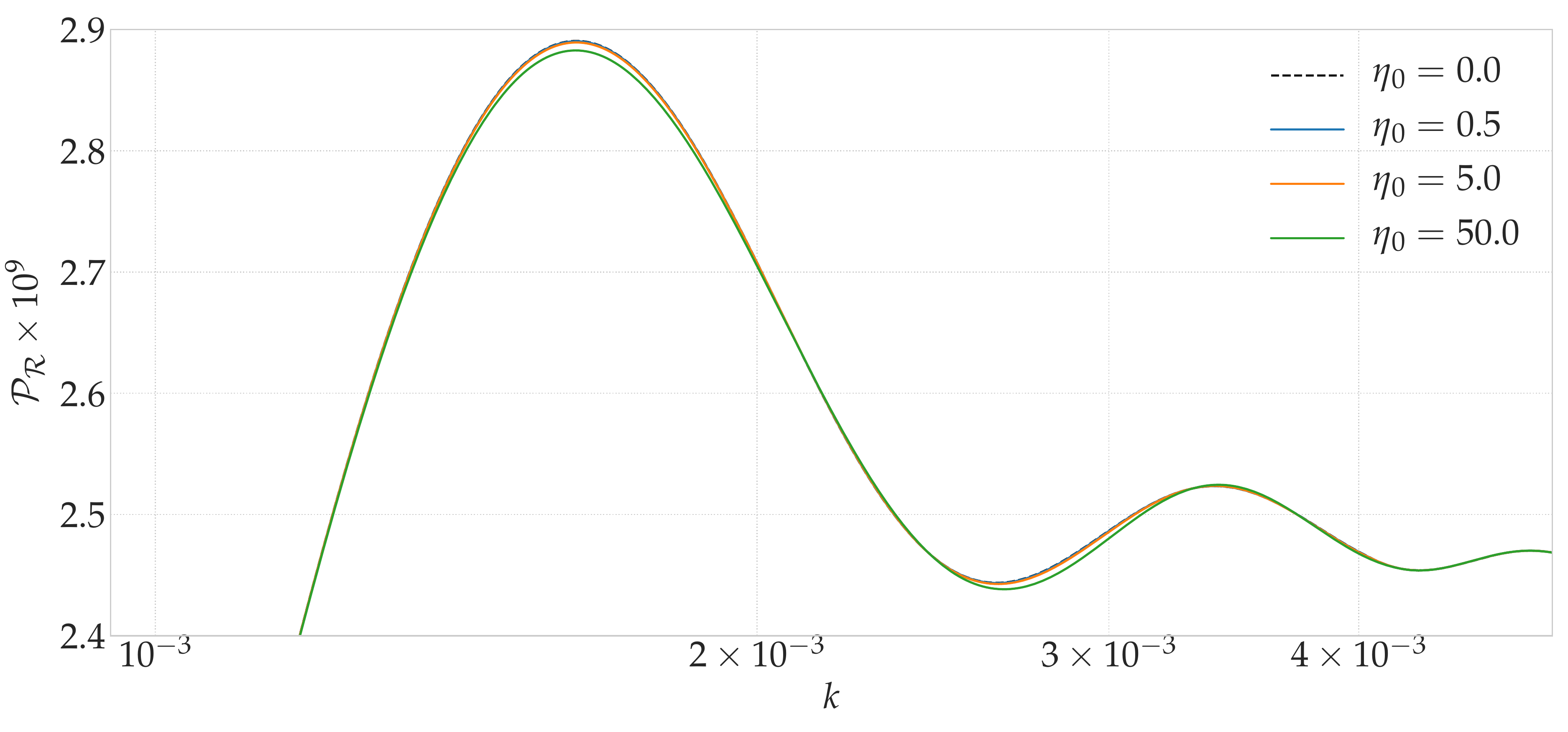}
		\caption{Primordial scalar power spectrum in hybrid LQC of the NO-vacuum defined with different choices of initial time between 0 (corresponding to the bounce) and 50, in Planck units. The background solution was studied in Section \ref{background}.}
		\label{fig11}
\end{figure}

On the other hand, the changes of initial time do not seem to have an important impact in the spectra, at least if $\eta_0$ remains well inside the kinetically dominated region. In Figure~\ref{fig11}, we show the changes in the primordial power spectrum for the NO-vacuum on the background solution of Section~\ref{background} when the initial time $\eta_0$ is varied. We see that the changes are small. We only display the interval $k\in[10^{-3},5\times 10^{-3}]$ because the changes are tiny outside this interval.
	
A different dependence of our results is on the parameters of the LQC quantization. There are two of these parameters, namely the Immirzi parameter $\gamma$ and the area gap $\Delta_g$. In our discussion so far, we have taken the standard values that are adopted for these parameters in LQC, based on arguments about black hole entropy and the spectrum of the area operator, respectively. We recall that these values are $\gamma=0.2375$ and $\Delta_g=4 \pi \sqrt{3}G \gamma $. The two parameters affect the value of the critical density, which~varies with them as $\rho_{\max}\propto \Delta_g^{-1} \gamma^{-2}\propto \gamma^{-3}$. Hence, when the Immirzi parameter increases, or~when the area gap increases (keeping $G$ constant), the critical density decreases. This~decrease makes the critical density approach the energy density where inflation begins (if the decrease is not radically large), reducing the region of kinetic dominance and placing the initial time where the vacuum is chosen closer to the inflationary period. Notice that if the decrease in the critical density were even larger, we would enter a completely different scenario where the bounce would be followed directly by a period of standard (potentially dominated) expansion. It would be interesting to study the modifications to the power spectra in this alternative scenario. On the other hand, let us point out that a change of the Immirzi parameter involves a rescaling of the physical volume, according to Equation \eqref{ABvariables}.  This rescaling changes the effective background solution. The new solution can be found with a matching process, integrating the effective trajectory for the new value of $\gamma$ backwards from a matching time that is sufficiently far away from the bounce, using there the rescaled value of the physical volume. This matching guarantees that the original and the new solutions share the same behavior for large volumes, since in that region the GR equations, that are independent of $\gamma$, are valid.
	
\begin{figure}[H]
		\centering
		\includegraphics[width=14 cm]{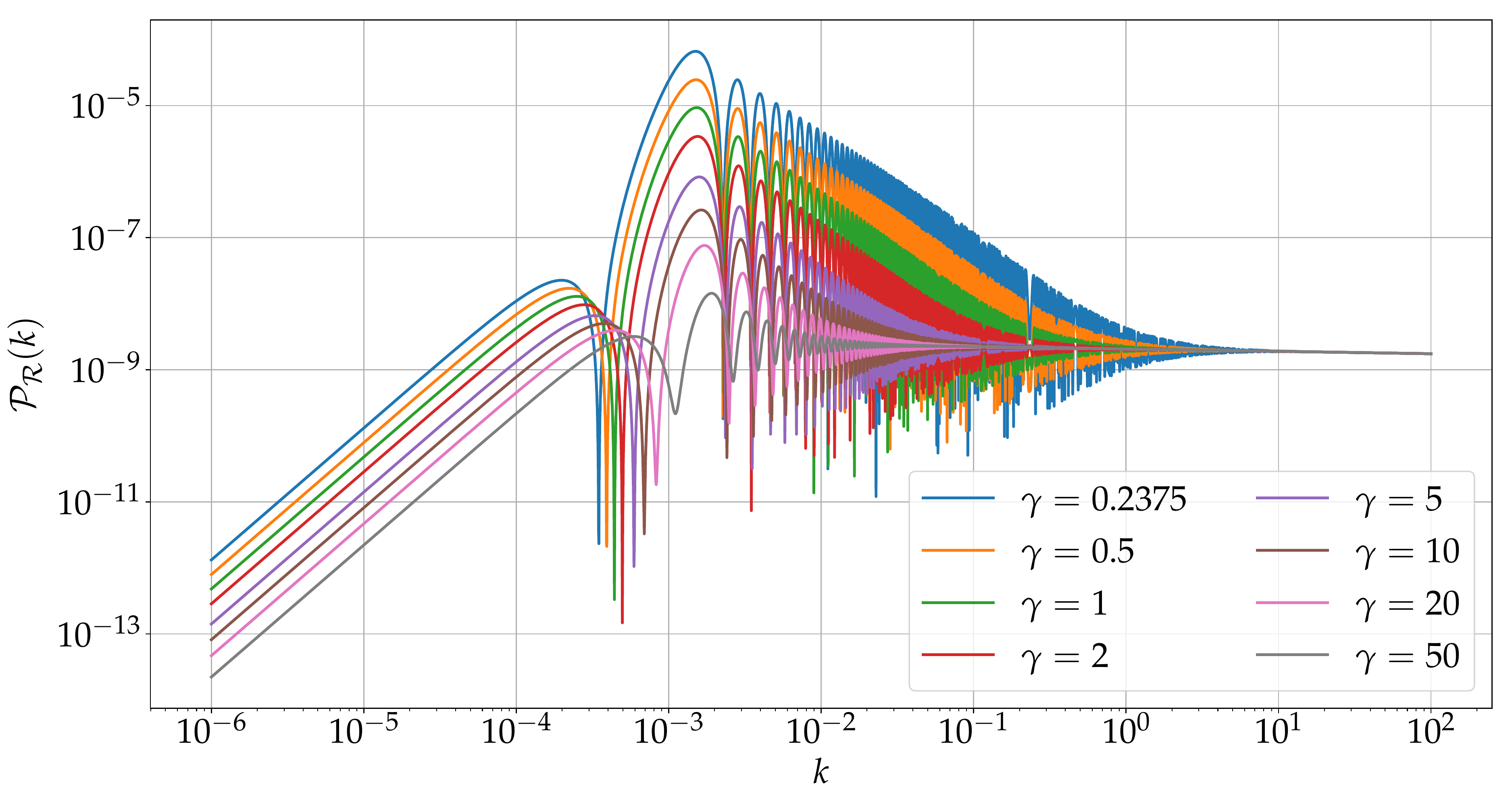}
		\caption{Primordial scalar power spectrum of the 0th-order adiabatic state in hybrid LQC for values of the Immirzi parameter between 0.2375 and 50. The background solution was studied in Section \ref{background}.}
		\label{fig12}
\end{figure}

\begin{figure}[H]
		\centering
		\includegraphics[width=14 cm]{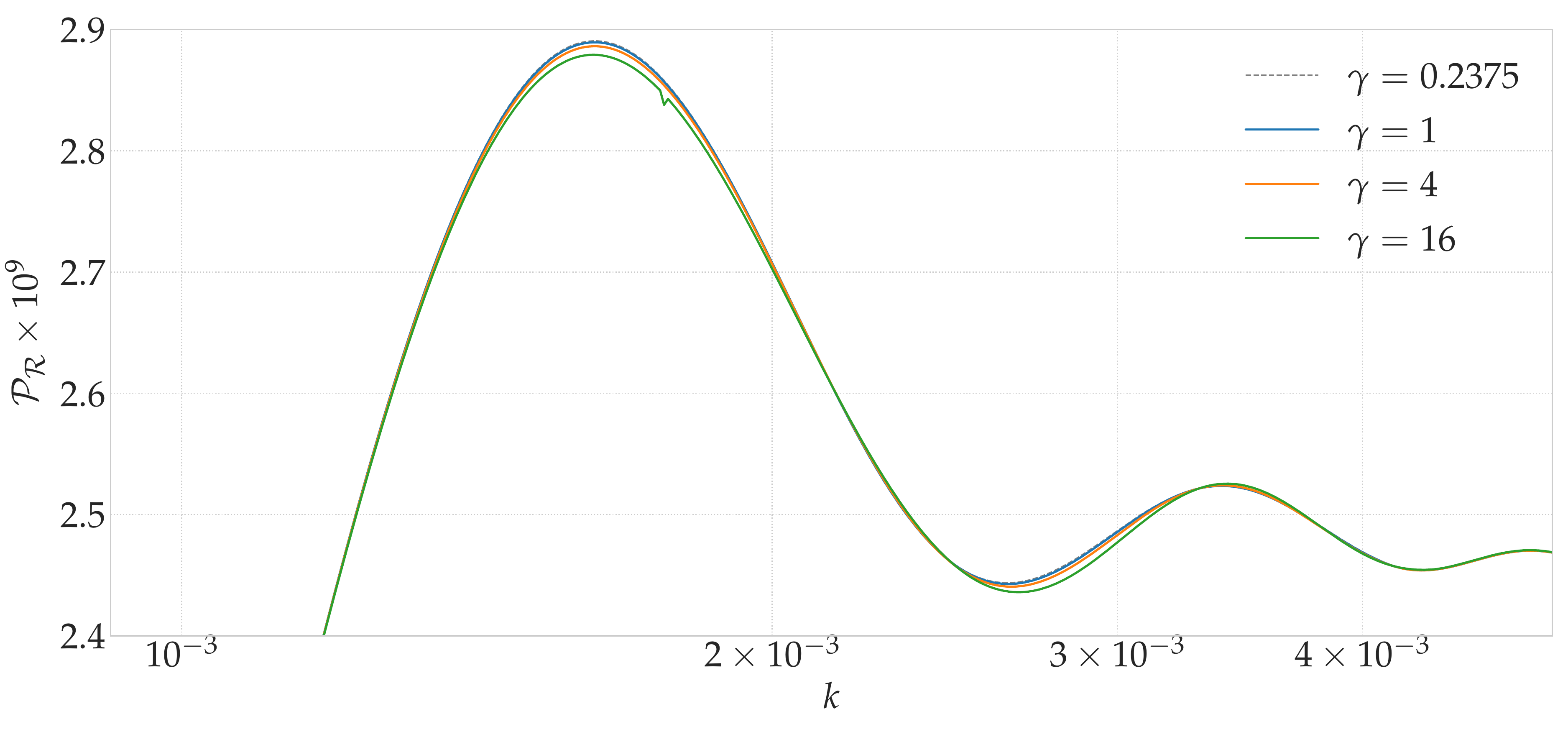}
		\caption{Primordial power spectrum of the NO-vacuum in hybrid LQC for values of the Immirzi parameter between 0.2375 and 16. The background solution was studied in Section \ref{background}.}
		\label{fig13}
\end{figure}

The primordial scalar power spectrum of the 0th-order adiabatic vacuum is displayed in Figure~\ref{fig12} for different values of the Immirzi parameter, including the standard value 0.2375. These spectra confirm our expectations that the oscillations get damped as $\gamma$ increases. For the NO-vacuum, on~the other hand, the primordial power spectrum is shown in Figure~\ref{fig13}. We plot only the interval  \mbox{$k\in[10^{-3},5\times 10^{-3}]$} because, again, the changes turn out to be negligible for other values of $k$. In~contrast with the situation encountered for the 0th-order adiabatic state, the variations in the spectrum are now very small. The background solution used in the computation of these spectra is, once more, the effective solution analyzed in Section \ref{background}.
	
\vspace{-12pt}  
	
\section{Conclusions}\label{conclusions}
	
We have discussed the role and determination of the vacuum state for quantum fields treated as perturbations in the framework of LQC. We have focused our attention on the hybrid approach to LQC, although most of our analysis can be extended to other proposals, like the dressed metric approach. For our discussion, we have considered an FLRW universe with compact flat topology in the presence of a massive scalar field, and we have perturbed it at quadratic order in the action. In~this perturbative formalism, the zero modes that describe the FLRW cosmology have been treated exactly at the adopted order of truncation. On the other hand, the non-compact case can be included in the study by taking an appropriate limit \cite{continuum}. In our system, formed by the homogeneous FLRW cosmology and its perturbations, we have introduced a canonical transformation that respects the gauge invariance at the perturbative level of truncation. With this transformation, the system can be described by MS and tensor gauge invariants, linear perturbative constraints and their momenta, and~zero modes. In a hybrid quantization, the physical states of this system depend only on the zero modes and on the perturbative gauge invariants. The sector of zero modes is quantized with the techniques of homogeneous LQC, while the perturbations are quantized with more standard, Fock~methods, using a Fock representation in a family of unitarily equivalent choices, that is selected by criteria based on invariance under spatial symmetries and unitarity of the Heisenberg evolution in the selected family of creation and annihilation operators. 
	
In this hybrid representation, the considered states are still subject to one constraint, provided by the zero mode of the total Hamiltonian. To search for solutions of physical interest, we~have adopted an ansatz of separation of variables, in which the states factorize in a product of partial wavefunctions, one for the homogeneous geometry and one for each of the modes of the gauge invariant perturbations. All of these partial wavefunctions are allowed to depend on the inflaton (the zero mode of the scalar field), that plays the role of an internal time in the system. Assuming that the perturbations mediate no relevant transitions of the FLRW geometry through the action of the constraint, and taking expectation values of this constraint in order to extract the most important information, one obtains a master constraint that governs the evolution of the perturbations. In~this way, one can derive dynamical equations for the gauge invariants, that can be written in a harmonic-oscillator form, with time-dependent masses that are supplied by (ratios of) expectation values of LQC operators in the quantum state that describes the FLRW geometry. We have then concentrated our attention on the case that this state is a(n approximate) solution of homogeneous LQC that remains peaked on a trajectory of the so-called effective dynamics. In this case, the time-dependent masses for the perturbations can be evaluated directly on these effective trajectories.
	
To integrate the evolution of the perturbations, as well as the effective dynamics of the homogeneous background, one needs to specify a set of initial conditions. These conditions are, e.g.,~the~value of the inflaton at the Big Bounce for the background, and the initial value of the gauge invariant modes and of their time derivatives for the perturbations. In addition, the background evolution depends on the value of the inflaton mass, that we regard as a parameter. We have found a sector of values of this mass and of the inflaton at the bounce such that the background solution leads to an inflationary era that can be compatible with the observations of the CMB, but still allows for imprints of quantum effects in the spectra, especially at large scales. We have analyzed the behavior of these kind of solutions, focusing the discussion on a specific background, corresponding to the choices $m=1.2\times 10^{-6}$ and $\phi_B=0.97$ in Planck units. In particular, we have seen that this solution presents a period of kinetic dominance, from the Big Bounce till the early stages of the inflationary process. 
	
Using this background solution, it is possible to integrate the equations of motion for the perturbations until the end of inflation, and then get the primordial power spectra needed to run the Boltzmann codes that provide at last the angular spectra that can be compared with the observational data. To carry out this program, we still need a piece of information of major relevance: the initial state of the perturbations. Fixing these initial data amounts to choosing a vacuum state at the initial time for the quantum fields of the gauge invariants. In this manner, we arrive at the most important issue studied in this work: the crucial dependence of the power spectra on the choice of vacuum (at least at large scales) and possible ways to fix this state, based on physical criteria.
	
Adiabatic states may be useful in order to constraint the ultraviolet behavior and obtain good regularization properties, but there is still an infinite freedom in the choice of an adiabatic vacuum. In~particular, we have considered two schemes to construct this type of states: obvious adiabatic states and iterative ones. In addition, in both schemes the state depends on the order of the asymptotic adiabatic approximation and on the initial time adopted for its construction. We have seen that the spectra of the iterative adiabatic states are most robust under changes of the adiabatic order than the obvious ones, and that they also provide a better behavior at the largest scales, with more suppression of power. Nonetheless, these states typically present a region with rapid oscillations in the spectrum and a consequent enhancement of power on average, something that seems to go against the behavior observed in the CMB. \footnote{After the completion of this work, we got notice of a recent work \cite{waco4} that constructs analytical solutions for the perturbations under certain approximations and employs them to define adiabatic states in the contracting phase of the background previous to the bounce, calculating the power spectrum and constraining the result by comparison with observations.}
	
We have discussed in detail an alternative proposal for the vacuum state: the non-oscillatory vacuum introduced by Mart\'{\i}n de Blas and Olmedo \cite{hybrpred}. The definition of the NO-vacuum state takes into account the dynamics of the perturbations in a time interval, and adjusts the vacuum to lead to a primordial power spectrum without large oscillations. We have seen that this vacuum provides a spectrum with a satisfactory suppression of power at large scales, like in the case of some adiabatic states, but in addition the resulting spectrum is now smooth and can fit very well the observations, making the proposal very appealing. In the case of the dressed metric, an apparently different vacuum that provides a spectrum with a good behavior and also fits the observations has been put forward by Ashtekar and Gupt \cite{AGvacio2}. It would be interesting to compute the corresponding spectrum for this vacuum in hybrid LQC and compare the results with the NO-vacuum.
	
To investigate the reasons behind the rapid oscillations and the averaged increase of power observed for the adiabatic states, we have considered initial conditions of the adiabatic type at later times, closer to the onset of inflation, as well as adiabatic states evolving with the dynamical equation of GR (on a background that is the GR counterpart of the effective solution studied in Section~\ref{background}). Our results indicate that these effects should be attributed to the choice of adiabatic data at an initial time that belongs to a kinetically dominated phase of the evolution, rather than to genuine quantum effects produced by the quantization of the geometry. The same oscillations and increase of averaged power appear also in GR. Actually, LQC puts a bound on the amplitude of those oscillations, because the energy density can never exceed the critical value $\rho_{\max}$ in the LQC scenario.
	
In addition, we have analyzed the sensitivity of the NO-vacuum spectrum to a change of the time interval employed for the definition of this vacuum. We have seen that the spectrum remains almost identical if the initial and final times are not changed considerably. Important modifications appear only if the interval does not cover the kinetically dominated epoch or if the final time of the interval approaches the bounce.
	
Finally, we have investigated the dependence of the power spectra on the value of the Immirzi parameter, assuming the hypothetical case that this parameter might be varied. The higher the Immirzi parameter gets, the lower the difference becomes between the energy density at the bounce and at the onset of inflation. This reduces the period of kinetic dominance, damping the oscillations in the case of adiabatic states. For the NO-vacuum, on the other hand, the spectrum turns out to be quite stable under a change of the Immirzi parameter.

\vspace{6pt}

\authorcontributions{Conceptualization: D.M.d.B.; B.E.N. and G.A.M.M. Original Draft Preparation: G.A.M.M. and B.E.N.;  Figures and Numerical Simulations: D.M.d.B.}

\funding{This work was supported by the Spanish MINECO grants number FIS2014-54800-C2-2-P and FIS2017-86497-C2-2-P.}
	
\acknowledgments{The authors are grateful to J. Olmedo for discussions.}
	
\conflictsofinterest{The authors declare no conflict of interest.} 
	
	
\reftitle{References}


\begin{thebibliography}{999}
\bibitem{precision1} Hinshaw, G.; Larson, D.; Komatsu, E.; Spergel, D.N.; Bennett, C.; Dunkley, J.; Nolta, M.R.; Halpern, M.; Hill, R.S.; Odegard, N.; et al. Nine-year Wilkinson Microwave Anisotropy Probe (WMAP) observations: Cosmological parameter results. {\em Astrophys. J. Suppl. Ser.} {\bf 2013}, {\em 208}, 19.
		
\bibitem{precision2} Ade, P.A.R. et al. [Planck Collaboration]. Planck 2015 results. XIII. Cosmological parameters. {\em  Astron. Astrophys.} {\bf 2016}, {\em 594}, A13.
		
		
\bibitem{planck-inf} Ade, P.A.R. et al. [Planck Collaboration]. Planck 2015 results. XX. Constraints on inflation.  {\em Astron. Astrophys.} {\bf 2016}, {\em 594}, A20.
		
\bibitem{planck1} Akrami, Y. et al. [Planck Collaboration]. Planck 2018 results. I. Overview and the cosmological legacy of Planck.  {\em arXiv} \textbf{2018}, arXiv:1807.06205.
		
\bibitem{planck2} Aghanim, N. et al. [Planck Collaboration]. Planck 2018 results. VI. Cosmological parameters. {\em arXiv} \textbf{2018}, arXiv:1807.06209.
		
\bibitem{waves1} Abbott, B.P. et al. [LIGO Scientific] and [Virgo Collaborations]. Observation of gravitational waves from a binary black hole merger. {\em Phys. Rev. Lett.} {\bf 2016}, {\em 116}, 061102. 
		
\bibitem{waves2} Abbott, B.P. et al. [LIGO Scientific] and [Virgo Collaborations]. Tests of General Relativity with GW150914. {\em Phys.~Rev.~Lett.} {\bf 2016}, {\em 116}, 221101. 
		
\bibitem{lqg} Thiemann, T. {\em Modern Canonical Quantum General Relativity}; Cambridge University Press: Cambridge, UK,~2007.
		
\bibitem{wald} Wald, R.M. {\em General Relativity}; University of Chicago Press: Chicago, IL, USA, 1984.
		
\bibitem{giulini} Giulini, D. Dynamical and Hamiltonian formulation of General Relativity. {\em arXiv} \textbf{2015}, arXiv:1505.01403.	
		
\bibitem{ap} Ashtekar, A.; Singh, P. Loop quantum cosmology: A status report. {\em Class. Quant. Grav.} {\bf 2011}, {\em 28}, 213001.
		
\bibitem{lqc1} Mena Marug\'an, G.A. A brief introduction to loop quantum cosmology. {\em AIP Conf. Proc.} {\bf 2011}, {\em 1130}, 89.
		
\bibitem{aps} Ashtekar, A.; Paw\l{}owski, T.; Singh, P. Quantum nature of the Big Bang: An analytical and numerical investigation. {\em Phys. Rev. D} {\bf 2006}, {\em 73}, 124038.
		
\bibitem{iaps} Ashtekar, A.; Paw\l{}owski, T.; Singh, P. Quantum nature of the big bang: Improved dynamics. {\em Phys. Rev. D} {\bf 2006}, {\em 74}, 084003.
		
\bibitem{mmo} Mart\'in-Benito, M.; Mena Marug\'an, G.A.; Olmedo, J. Further improvements in the understanding of isotropic loop quantum cosmology. {\em Phys. Rev. D} {\bf 2009}, {\em 80}, 104015.
		
\bibitem{taveras} Taveras, V. LQC corrections to the Friedmann equations for a universe with a free scalar field. {\em Phys. Rev. D} {\bf 2008}, {\em 78}, 064072.
		
\bibitem{ACH} Ashtekar, A.; Campiglia, M.; Henderson, A. Path integrals and the WKB approximation in loop quantum cosmology. {\em Phys. Rev. D} {\bf 2010}, {\em 82}, 124043.
		
\bibitem{effective0} Wilson-Ewing, E. Holonomy corrections in the effective equations for scalar mode perturbations in loop quantum cosmology. {\em Class. Quant. Grav.} {\bf 2012}, {\em 29}, 085005.
		
\bibitem{effective} Cailleteau, T.; Linsefors, L.; Barreau, A. Anomaly-free perturbations with inverse-volume and holonomy corrections in loop quantum cosmology. {\em Class. Quant. Grav.} {\bf 2014}, {\em 31}, 125011.
		
\bibitem{hybr1} Fern\'andez-M\'endez, M.; Mena Marug\'an, G.A.; Olmedo, J. Hybrid quantization of an inflationary universe. {\em Phys. Rev. D} {\bf 2012}, {\em 86}, 024003.
		
\bibitem{dressed1} Agullo, I.; Ashtekar, A.; Nelson, W. A quantum gravity extension of the inflationary scenario. {\em Phys. Rev. Lett.} {\bf 2012}, {\em 109}, 251301.
		
\bibitem{edward} Wilson-Ewing, E. Testing loop quantum cosmology. {\em Comptes Rendus Phys.} {\bf 2017}, {\em 18}, 207--225.
		
\bibitem{hybr2} Fern\'andez-M\'endez, M.; Mena Marug\'an, G.A.; Olmedo, J. Hybrid quantization of an inflationary model: The~flat case. {\em Phys. Rev. D} {\bf 2013}, {\em 88}, 044013.
		
\bibitem{hybrnum} Fern\'andez-M\'endez, M.; Mena Marug\'an, G.A.; Olmedo, J. Effective dynamics of scalar perturbations in a flat Friedmann-Robertson-Walker spacetime in loop quantum cosmology. {\em Phys. Rev. D} {\bf 2014}, {\em 89},  044041.
		
\bibitem{hybr3} Castell\'o Gomar, L.; Fern\'andez-M\'endez, M.; Mena Marug\'an, G.A.; Olmedo, J. Cosmological perturbations in hybrid loop quantum cosmology: Mukhanov--Sasaki variables. {\em Phys. Rev. D} {\bf 2014}, {\em 90}, 064015.
		
\bibitem{hybr4} Castell\'o Gomar, L.; Mart\'{\i}n-Benito, M.; Mena Marug\'an, G.A. Gauge-invariant perturbations in hybrid quantum cosmology. {\em J. Cosmol. Astropart. Phys.} {\bf 2015}, {\em 2015}, 045. 
		
\bibitem{hybr5} Castell\'o Gomar, L.; Mart\'{\i}n-Benito, M.; Mena Marug\'an, G.A. Quantum corrections to the Mukhanov-Sasaki equations. {\em Phys. Rev. D} {\bf 2016}, {\em 93}, 104025.
		
\bibitem{hybrten} Ben\'itez Mart\'inez, F.; Olmedo, J. Primordial tensor modes of the early universe. {\em Phys. Rev. D} {\bf 2016}, {\em 93}, 124008.
		
\bibitem{hybrpred} Mart\'in de Blas, D.; Olmedo, J. Primordial power spectra for scalar perturbations in loop quantum cosmology. {\em J. Cosmol. Astropart. Phys.} {\bf 2016}, {\em 2016}, 029. 
		
\bibitem{hybrgui} Castell\'o Gomar, L.; Mart\'in de Blas, D.; Mena Marug\'an, G.A.; Olmedo, J. Hybrid loop quantum cosmology and predictions for the cosmic microwave background. {\em Phys. Rev. D} {\bf 2017}, {\em 96}, 103528.
		
\bibitem{hybrferm} Elizaga Navascu\'es, B.; Mart\'{\i}n-Benito, M.; Mena Marug\'an, G.A. Fermions in hybrid loop quantum cosmology. {\em Phys. Rev. D} {\bf 2017}, {\em 96}, 044023.
		
\bibitem{dressed2} Agullo, I.; Ashtekar, A.; Nelson, W. Extension of the quantum theory of cosmological perturbations to the Planck era. {\em Phys. Rev. D} {\bf 2013}, {\em 87}, 043507.
		
\bibitem{dressed3} Agullo, I.; Ashtekar, A.; Nelson, W. The pre-inflationary dynamics of loop quantum cosmology: Confronting quantum gravity with observations. {\em Class. Quant. Grav.} {\bf 2013}, {\em 30}, 085014.
		
\bibitem{AMorris} Agullo, I.; Morris, N.A. Detailed analysis of the predictions of loop quantum cosmology for the primordial power spectra. {\em Phys. Rev. D} {\bf 2015}, {\em 92}, 124040.
		
\bibitem{Agullo} Agullo, I. Loop quantum cosmology, non-Gaussianity, and CMB power asymmetry. {\em Phys. Rev. D} {\bf 2015}, {\em 92},~064038.
		
\bibitem{AAG} Agullo, I.; Ashtekar, A.; Gupt, B. Phenomenology with fluctuating quantum geometries in loop quantum cosmology. {\em Class. Quant. Grav.} {\bf 2017}, {\em 34}, 074003.
		
\bibitem{ABS} Agullo, I.; Bolliet, B.; Sreenath, V. Non-Gaussianity in loop quantum cosmology. {\em Phys. Rev. D} {\bf 2018}, {\em 97},~066021.
		
\bibitem{hybrmass} Elizaga Navascu\'es, B; Mart\'{\i}n de Blas, D.; Mena Marug\'an, G.A. Time-dependent mass of cosmological perturbations in the hybrid and dressed metric approaches to loop quantum cosmology. {\em Phys. Rev. D} {\bf 2018}, {\em 97}, 043523.
		
\bibitem{hybrgowdy1} Mart\'{\i}n-Benito, M.; Garay, L.J.; Mena Marug\'{a}n, G.A. Hybrid quantum Gowdy cosmology: Combining loop and Fock quantizations. {\em Phys. Rev. D} {\bf 2008}, {\em 78}, 083516.
		
\bibitem{hybrgowdy2} Mena Marug\'{a}n G.A.; Mart\'{\i}n-Benito, M. Hybrid quantum cosmology: Combining loop and Fock quantizations. {\em Int. J. Mod. Phys. A} {\bf 2009}, {\em 24}, 2820--2838.
		
\bibitem{hybrgowdy3} Mart\'{\i}n-Benito, M.; Mena Marug\'{a}n, G.A.; Wilson-Ewing, E. Hybrid quantization: From Bianchi I to the Gowdy model. {\em Phys. Rev. D} {\bf 2010}, {\em 82}, 084012.
		
\bibitem{hybrgowdymatt} Mart\'{\i}n-Benito, M.; Mart\'{\i}n-de Blas, D.; Mena Marug\'{a}n, G.A. Matter in inhomogeneous loop quantum cosmology: The Gowdy T3 model. {\em Phys. Rev. D} {\bf 2011}, {\em 83}, 084050.
		
\bibitem{hybrid} Elizaga Navascu\'es, B.; Mart\'{\i}n-Benito, M.; Mena Marug\'an, G.A. Hybrid models in loop quantum cosmology. {\em Int. J. Mod. Phys. D} {\bf 2016}, {\em 25}, 1642007.
		
\bibitem{uniqueness1} Fern\'andez-M\'endez, M.; Mena Marug\'an, G.A.; Olmedo, J.; Velhinho, J.M. Unique Fock quantization of scalar cosmological perturbations. {\em Phys. Rev. D} {\bf 2012}, {\em 85}, 103525.
		
\bibitem{uniquenessferm} Cortez, J.; Elizaga Navascu\'es, B.; Mart\'in-Benito, M.; Mena Marug\'an, G.A.; Velhinho, J.M. Dirac fields in flat FLRW cosmology: Uniqueness of the Fock quantization. {\em Ann. Phys.} {\bf 2017}, {\em 376}, 76--88.
		
\bibitem{unitarity1} Cortez, J.;  Mena Marug\'an, G.A.; Olmedo, J.; Velhinho, J.M. A uniqueness criterion for the Fock quantization of scalar fields with time-dependent mass. {\em Class. Quant. Grav.} {\bf 2011}, {\em 28}, 172001.
		
\bibitem{unitarity2} Cortez, J.; Mena Marug\'an, G.A.; Olmedo, J.; Velhinho, J.M. Criteria for the determination of time dependent scalings in the Fock quantization of scalar fields with a time dependent mass in ultrastatic spacetimes. {\em Phys.~Rev.~D} {\bf 2012}, {\em 86}, 104003.
		
\bibitem{unitarity3} Castell\'o Gomar, L.; Cortez, J.; Mart\'in-de Blas, D.; Mena Marug\'an, G.A.; Velhinho, J.M. Uniqueness of the Fock quantization of scalar fields in spatially flat cosmological spacetimes. {\em J. Cosmol. Astropart. Phys.} {\bf 2012}, {\em 2012}, 001.   

		
\bibitem{modeunit} Cortez, J.; Fonseca, L.; Mart\'{\i}n-de Blas, D.; Mena Marug\'an, G.A. Uniqueness of the Fock quantization of scalar fields under mode preserving canonical transformations varying in time. {\em Phys. Rev. D} {\bf 2013}, {\em 87}, 044013.
		
\bibitem{unitarity4} Cortez, J.; Mart\'{\i}n-de Blas, D.; Mena Marug\'an, G.A.; Velhinho, J.M. Massless scalar field in de Sitter spacetime: Unitary quantum time evolution. {\em Class. Quant. Grav.} {\bf 2013}, {\em 30}, 075015.
		
\bibitem{unitarity5} Cortez, J.; Mena Marug\'an, G.A.; Velhinho, J.M. Quantum unitary dynamics in cosmological spacetimes. {\em Ann.~Phys.} {\bf 2015}, {\em 363}, 36--47.
		
\bibitem{bianchi} Cortez, J.; Elizaga Navascu\'es, B.; Mart\'in-Benito, M.; Mena Marug\'an, G.A.; Olmedo, J.; Velhinho, J.M. Uniqueness of the Fock quantization of scalar fields in a Bianchi I cosmology with unitary dynamics. {\em Phys.~Rev.~D} {\bf 2016}, {\em 94}, 105019.
		
\bibitem{hybrfermback} Elizaga Navascués, B.; Mena Marugán, G.A.; Prado Loy, S. Backreaction of fermionic perturbations in the Hamiltonian of hybrid loop quantum cosmology. {\em arXiv} \textbf{2018}, arXiv:1805.04133.
		
\bibitem{immirzi} Immirzi, G. Quantum gravity and Regge calculus. {\em Nucl. Phys. Proc. Suppl.} {\bf 1997}, {\em 57}, 65--72.
		
\bibitem{BHlqg1} Meissner, K.A. Black-hole entropy in loop quantum gravity. {\em Class. Quant. Grav.} {\bf 2004}, {\em 21}, 5245.
		
\bibitem{BHlqg2} Domagala, M.; Lewandowski, J. Black-hole entropy from quantum geometry. {\em Class. Quant. Grav.} {\bf 2004}, {\em 21},~5233.
		
\bibitem{bardeen} Bardeen, J.M. Gauge-invariant cosmological perturbations. {\em Phys. Rev. D} {\bf 1980}, {\em 22}, 1882--1905. 
		
\bibitem{mukhanov} Mukhanov, V. Quantum theory of gauge-invariant cosmological perturbations. {\em Zh. Eksp. Teor. Fiz.} {\bf 1988}, {\em 94}, 1--11. 
		
\bibitem{sasaki} Sasaki, M. Gauge invariant scalar perturbations in the new inflationary universe. {\em Prog. Theor. Phys.} {\bf 1983}, {\em 70}, 394--411.
		
\bibitem{kodasasa} Kodama, H.; Sasaki, M. Cosmological perturbation theory. {\em Prog. Theor. Phys. Suppl.} {\bf 1984}, {\em 78}, 1--166.
		
\bibitem{continuum} Elizaga Navascu\'es, B.; Mena Marug\'an, G.A. Perturbations in hybrid loop quantum cosmology: Continuum limit in Fourier space. {\em arXiv} \textbf{2018}, arXiv:1809.03884.
		
\bibitem{AGvacio1} Ashtekar, A.; Gupt, B. Initial conditions for cosmological perturbations. {\em Class. Quant. Grav.} {\bf 2017}, {\em 34}, 035004.
		
\bibitem{adiabatic} Parker, L. Quantized fields and particle creation in expanding universes. I. {\em Phys. Rev.} {\bf 1969}, {\em 183}, 1057.
		
\bibitem{adiabaticLR} L\"uders, C.; Roberts, J.E. Local quasiequivalence and adiabatic vacuum states. {\em Commun. Math. Phys.} {\bf 1990}, {\em 134}, 29--63.
		
\bibitem{adiabaticreg1} Parker, L.; Fulling, S.A. Adiabatic regularization of the energy-momentum tensor of a quantized field in homogeneous spaces. {\em Phys. Rev. D} {\bf 1974}, {\em 9}, 341--354.
		
\bibitem{adiabaticreg2} Anderson, P.R.; Parker, L. Adiabatic regularization in closed Robertson-Walker universes. {\em Phys. Rev. D} {\bf 1987}, {\em 36}, 2963--2969.
		
\bibitem{AANvacio} Agullo, I.; Nelson, W.; Ashtekar, A. Preferred instantaneous vacuum for linear scalar fields in cosmological space-times. {\em Phys. Rev. D} {\bf 2015}, {\em 91}, 064051.
		
\bibitem{AGvacio2} Ashtekar, A.; Gupt, B. Quantum gravity in the sky: Interplay between fundamental theory and observations. {\em Class. Quant. Grav.} {\bf 2017}, {\em 34}, 014002.
		
\bibitem{adinv} Bertoni, C.; Finelli, F.; Venturi, G. Adiabatic invariants and scalar fields in a de Sitter space-time. {\em Phys. Lett. A} {\bf 1998}, {\em 237}, 331--336.
		
\bibitem{slowroll} Liddle, A.R.; Lyth, D.H. {\em Cosmological Inflation and Large-Scale Structure}; Cambridge University Press: Cambridge, UK, 2000.
		
\bibitem{efolds} Liddle, A.R.; Leach, S.M. How long before the end of inflation were observable perturbations produced? {\em Phys. Rev. D} {\bf 2003}, {\em 68}, 103503.
		
\bibitem{linde} Contaldi, C.R.; Peloso, M.; Kofman, L.; Linde, A. Suppressing the lower multipoles in the CMB anisotropies. {\em  J. Cosmol. Astropart. Phys.} {\bf 2003}, {\em 2003}, 196.
		
\bibitem{waco1} Zhu, T.; Wang, A.; Kirsten, K.; Cleaver, G.; Sheng, Q. Universal features of quantum bounce in loop quantum cosmology. {\em Phys. Lett. B} {\bf 2017}, {\em 773}, 196--202.
		
\bibitem{waco2} Zhu, T.; Wang, A.; Kirsten, K.; Cleaver, G.; Sheng, Q. Pre-inflationary universe in loop quantum cosmology. {\em Phys. Rev. D} {\bf 2017}, {\em 96}, 083520.
		
\bibitem{waco3} Shahalam, M.; Sharma, M.; Wu, Q.; Wang, A. Pre-inflationary dynamics in loop quantum cosmology: Power-law potentials. {\em Phys. Rev. D} {\bf 2017}, {\em 96}, 123533.		
		
\bibitem{class} Blas, D.; Lesgourgues, J.; Tram, T. The Cosmic Linear Anisotropy Solving System (CLASS) Part II: Approximation schemes. {\em  J. Cosmol. Astropart. Phys.} {\bf 2011}, {\em 2011}, 034.
		
\bibitem{waco4} Wu, Q.; Zhu, T.; Wang, A.  Non-adiabatic evolution of primordial perturbations and non-gaussianity in hybrid approach of loop quantum cosmology. {\em arXiv} \textbf{2018}, arXiv:1809.03172.
		
\end{thebibliography}
\end{document}